\documentclass[journal]{IEEEtran}

\usepackage{threeparttable}
\usepackage{cite}
\usepackage{mathrsfs}
\usepackage{bbm}
\usepackage{amssymb}
\usepackage{bbding}
\usepackage{threeparttable}
\usepackage{graphicx}
\usepackage[mathcal]{euscript}

\usepackage{psfrag,calc,url,bm}
\usepackage{diagbox}
\usepackage{cite}

\usepackage{graphicx}
\usepackage{siunitx}
\usepackage{psfrag}

\usepackage{subfigure}
\usepackage{hyperref}
\usepackage{url}

\usepackage{stfloats}

\usepackage{amsmath}

\usepackage{float}

\usepackage{algorithmic}

\usepackage[ruled,linesnumbered,vlined]{algorithm2e}

\usepackage{color}
\usepackage{makecell}
\usepackage{boxedminipage}

\usepackage{amsthm}

\usepackage{multirow}

\usepackage{setspace}

\usepackage{soul}
\usepackage{epstopdf}

\usepackage{booktabs}
\allowdisplaybreaks[3]

\begin{document}

\title{Two-Stage Heterogeneous Graph Neural Network for RIS-Aided Physical-Layer Security}
\author{Zihan Song, Yang Lu,~\IEEEmembership{Senior Member,~IEEE}, Wei Chen,~\IEEEmembership{Senior Member,~IEEE}, Bo Ai,~\IEEEmembership{Fellow,~IEEE}, \\Zhiguo  Ding,~\IEEEmembership{Fellow,~IEEE}, and Arumugam Nallanathan,~\IEEEmembership{Fellow,~IEEE}
\thanks{Zihan Song and Yang Lu are with the School of Computer and Information Technology, Beijing Jiaotong University, Beijing 100044, China (e-mail: 23120405@bjtu.edu.cn, yanglu@bjtu.edu.cn).}
\thanks{Wei Chen and Bo Ai are with the School of Electronics and Information Engineering, Beijing Jiaotong University, Beijing 100044, China (e-mail: weich@bjtu.edu.cn, boai@bjtu.edu.cn).}
\thanks{Zhiguo Ding is with the School of Electrical and Electronic Engineering (EEE), Nanyang Technological University, Singapore 639798 (Zhiguo.ding@ntu.edu.sg).}
\thanks{Arumugam Nallanathan is with the School of Electronic Engineering and Computer Science, Queen Mary University of London, London and also with the Department of Electronic Engineering, Kyung Hee University, Yongin-si, Gyeonggi-do 17104, South Korea (e-mail: a.nallanathan@qmul.ac.uk).}
%\thanks{The code is available at xxx.}
}

\maketitle

\begin{abstract}
This paper investigates physical-layer security (PLS) enabled by graph neural networks (GNNs). We propose a two-stage heterogeneous GNN (HGNN) to maximize the secrecy energy efficiency (SEE) of a reconfigurable intelligent surface (RIS)-assisted multi-input-single-output (MISO) system that serves multiple legitimate users (LUs) and eavesdroppers (Eves). The first stage formulates the system as a bipartite graph involving three types of nodes—RIS reflecting elements, LUs, and Eves—with the goal of generating the RIS phase shift matrix. The second stage models the system as a fully connected graph with two types of nodes (LUs and Eves), aiming to produce beamforming and artificial noise (AN) vectors. Both stages adopt an HGNN integrated with a multi-head attention mechanism, and the second stage incorporates two output methods: beam-direct and model-based approaches. The two-stage HGNN is trained in an unsupervised manner and designed to scale with the number of RIS reflecting elements, LUs, and Eves. Numerical results demonstrate that the proposed two-stage HGNN outperforms state-of-the-art GNNs in RIS-aided PLS scenarios. Compared with convex optimization algorithms, it reduces the average running time by three orders of magnitude with a performance loss of less than $4\%$. Additionally, the scalability of the two-stage HGNN is validated through extensive simulations.
\end{abstract}

\section{Introduction}

\subsection{Background}
RIS has been widely regarded as a key wireless coverage enabler for next-generation mobile communication systems \cite{bck1}. Specifically, RIS can intelligently adjust the phase of incident electromagnetic waves without additional radio frequency chains, thus effectively improving wireless channel quality, extending coverage, and mitigating signal interference \cite{bck2}. Its passive operation enables energy-efficient deployment, which aligns well with the requirements of green communications. The channel reconfigurability of RIS can enlarge the channel diversity among receivers, which can be leveraged to enhance PLS \cite{bck3}. As demonstrated in \cite{hong_tcom_2020},  RIS-aided PLS can significantly improve the system SEE and secrecy sum rate (SSR) even when there are no direct communication links, and the performance gain increases with the number of reflecting elements. Despite the advantages of RIS, efficiently optimizing RIS-assisted communication systems is non-trivial due to the deeply coupled BS-related and RIS-related parameters.

Existing optimization approaches for RIS can be categorized into three categories: 1) convex optimization (CVXopt)-based approach, 2) deep reinforcement learning (DRL)-based approach, and 3) learn-to-optimize approach. For example, the authors in \cite{Amiriara2023IRS-User}, \cite{lu2023secrecy} and \cite{Guo_twc_2020} investigated multi-user MISO systems assisted by RIS, employing CVXopt-based approaches to maximize the system SEE and sum rate, respectively. However, as reviewed in \cite{pan_jstsp_2022}, most existing CVXopt-based methods mainly rely on alternating optimization (AO) or block coordinate descent (BCD) to tackle the deeply coupled variables, leading to high computational overhead and separate optimization losses. In contrast, DRL-based approaches have emerged as a promising joint-optimization solver for RIS-assisted systems. They have been validated to enable energy-efficient designs in RIS-assisted MISO \cite{huang_jsac_2020}, simultaneous wireless information and power transfer   \cite{Zhang_jsac_2023}, and unmanned aerial vehicle mobile edge computing systems \cite{Abdalla_ICC_2025}.  Nevertheless, DRL-based approaches learn from interactions in partially controllable wireless environments, thus suffering from limited generalization capability.

This paper focuses on the data-driven learn-to-optimize approach, which has been widely adopted as an efficient optimization solver for various wireless systems \cite{sun_tsp_2018}. Specifically,
existing studies have adopted multi-layer perceptrons (MLPs)  \cite{Saikia2023Beamforming,Gao2020Unsupervised} and convolutional neural network (CNNs) \cite{Ma2026Unsupervised,zhang2023Passive}, respectively, to realize joint   optimization for RIS-assisted  systems. However, MLPs and CNNs are inherently limited by their fixed input and output dimensions, thus failing to provide satisfactory generalization performance in dynamic RIS-assisted systems. By contrast, GNNs can effectively overcome the limitation \cite{gnn1,gnn2}. As analyzed in \cite{shen_twc_2023}, GNNs are well matched to the topology of wireless networks, thereby not only outperforming other models but also achieving strong adaptability to wireless dynamics. 
For example, GNNs have been shown to enable real-time, near-optimal and scalable inference for MISO systems, aiming to maximize the sum rate (SR) \cite{li_twc_2024}, energy efficiency (EE) \cite{he2025ICGNN} and SEE \cite{Song_tvt_2024}, respectively. Compared with non-RIS and non-PLS systems, RIS-aided PLS involves complicated heterogeneity among distinct entities, such as the BS, RIS, LUs, and Eves. Therefore, it is of great importance to align the deep learning (DL) model with both the system and the task. 
Existing works handle the ``heterogeneity" via two perspectives, i.e., framework and neural architecture. For the framework, the authors in \cite{Yang2024GATPrecoding,he2025ris} developed a multi-stage GNN to sequentially generate the phase shift matrix and beamforming vectors. This framework inherits the algorithmic structure of conventional CVXopt-based approaches and enhances the model’s interpretability. For the neural architecture, the authors in \cite{yeh2024enhanced,tang2025joint, Yin2026Graph,liu2025beamforming} exploited  HGNNs to capture the heterogeneity within RIS-assisted systems. These HGNNs exhibit superior scalability with respect to the system entities.

\subsection{Contribution}
Existing  works have demonstrated that the model's expressive capability can be enhanced  from the perspectives of both framework and neural architecture. However, no existing work has yet integrated the advantages of these two aspects. To fill this gap, this paper proposes to utilize the two-stage framework and HGNNs for RIS-aided physical-layer security, which involves heterogeneous system entities (i.e., BS, RIS, LUs, and Eves) as well as dual-functional services. The contributions of this paper are summarized as follows.
\begin{itemize}
    \item We consider a typical RIS-aided PLS system, in which a multi-antenna BS serves multiple LUs in the presence of Eves via both direct and cascaded links. The transmitted signals consist of both information-bearing signals for LUs and AN to disrupt Eves. We formulate an SEE maximization problem by jointly optimizing the beamforming and AN vectors, as well as the RIS phase shift matrix.
    \item We propose a two-stage HGNN to solve the formulated problem. Specifically, the first stage models the considered system as a bipartite graph and employs both edge-based and edge-free graph attention operators to map channel state information (CSI) to the phase shift matrix. The second stage models the  system as a fully-connected  graph with heterogeneous node types and adopts an edge-free graph attention operator and a semantic attention mechanism to map augmented effective CSI to the beamforming and AN vectors. The outputs of both stages are fed into an unsupervised loss function for end-to-end joint training. 
    
    \item  Analysis results are presented to validate that the proposed two-stage HGNN is scalable with respect to the number of RIS reflecting elements, LUs, and Eves. Numerical results are provided to corroborate the analytical findings. Furthermore, the superiority of the proposed two-stage HGNN is demonstrated through comparisons with the state-of-the-art GNNs. Additionally, ablation experiments are conducted to validate the effectiveness of the key components of the proposed model.  %proposed two-stage framework can effectively reduce the fitting complexity of the target problem and achieve better performance.
\end{itemize}

The rest of this paper is organized as follows. Section II reviews emerging AI-based RIS-aided wireless communication technologies. Section III introduces the system model and problem formulation. Section IV presents two heterogeneous graph representation methods for the system. Section V details the proposed two-stage HGNN, followed by a scalability analysis. Section VI presents numerical results and evaluations. Section VII concludes the paper.

{{\it Notation}: The following mathematical notations and symbols are used throughout this paper. Bold lowercase letters (e.g., $\mathbf{a}$) denote column vectors, and bold uppercase letters (e.g., $\mathbf{A}$) denote matrices or higher-dimensional tensors. The set of real numbers is denoted by $\mathbb{R}$, and the set of $n \times m$ real matrices by $\mathbb{R}^{n \times m}$. Similarly, $\mathbb{C}^n$ and $\mathbb{C}^{n \times m}$ denote the sets of $n$-dimensional complex column vectors and $n \times m$ complex matrices, respectively. For a complex number $a$, $|a|$ denotes its modulus, and $\Re(a)$ and $\Im(a)$ denote its real and imaginary parts, respectively. For a vector $\mathbf{a}$, $\| \mathbf{a} \|$ denotes its Euclidean norm. For a matrix $\mathbf{A}$, $\mathbf{A}^T$, $\mathbf{A}^H$, $\| \mathbf{A} \|$, and $\text{Trace}(\mathbf{A})$ represent its transpose, conjugate transpose, Frobenius norm, and trace (i.e., the sum of its diagonal elements), respectively. The notation $\mathbf{a}[i]$ denotes the $i$-th entry of vector $\mathbf{a}$; $\mathbf{A}[i,j]$ denotes the element in the $i$-th row and $j$-th column of matrix $\mathbf{A}$; and $\mathbf{A}[i,:]$ denotes the entire $i$-th row of $\mathbf{A}$. The operator $\text{Concat}(\cdot)$ denotes the concatenation of its inputs. The operator $\text{LeakyReLU}(\cdot)$ denotes the leaky rectified linear unit (ReLU) with a small negative slope for negative inputs.

\section{Related Works}

This section summarizes the related works on GNN-enabled RIS-assisted systems. To clearly highlight the novelty of this paper, we compare the proposed two-stage HGNN and existing GNN-based models in Table \ref{com}.

\subsection{GNN-Enabled RIS Transmission}

Various GNNs have been leveraged to jointly optimize RIS-assisted systems. The authors in \cite{Yang2024GATPrecoding} modeled an RIS-assisted MISO system as a homogeneous graph and proposed a two-stage framework to solve a sum-rate maximization problem, where the first stage employed a graph attention network (GAT) to predict the phase shift matrix and the second stage utilized an MLP to generate beamforming vectors. The authors in \cite{he2025ris} considered an RIS-assisted downlink pinching-antenna system, with the objective of optimizing both the sum rate (SR) and energy efficiency (EE). They proposed a three-stage GNN, where each stage modeled the system as a fully connected graph and adopted either a graph convolutional network (GCN)\cite{GCN} or a GAT \cite{GAT} to derive the transmit parameters.

In addition to the homogeneous graph, the heterogeneous graphs may be more suitable for RIS-assisted systems and enable the problem to be solved within a single-stage framework.  The authors in \cite{yeh2024enhanced} formulated an RIS-assisted MISO mmWave system into a heterogeneous graph with an RIS node and multiple user nodes, and proposed a modified GAT to map angular cascaded channels to beamforming vectors and reflection coefficients for sum-rate maximization. Besides, they adopted transfer learning in the model to adapt to variations in user counts. The authors in \cite{tang2025joint} formulated a sum-rate maximization problem for a system utilizing collaboration of an RIS and a relay, which includes two phases, i.e.,  BS-RIS-user phase and relay-RIS-user phase. Regarding the first phase, they modeled the system as a fully-connected heterogeneous graph with one RIS node and multiple user nodes, and adopted an HGNN to yield phase shift matrix and beamforming vectors via different message aggregation methods. The authors in \cite{Yin2026Graph} considered an RIS-assisted MISO system covering both single-cell and cell-free scenarios and formulated an EE maximization problem. The system was represented as a heterogeneous graph with one RIS node and multiple user nodes, which was fed into a GNN with two distinct feature initialization strategies to jointly optimize  and the phase shift matrices and  beamforming vectors.  The authors in \cite{liu2025beamforming} considered a multi-RIS-assisted system and formulated a sum-rate maximization problem. The system is represented by a fully-connected heterogeneous graph with a BS node, multiple RIS nodes and user nodes,  and fed into an HGNN to jointly obtain the phase shift matrix,  beamforming vectors and the RIS association matrix.

However, since all reflecting elements are represented by a single RIS node, the models in \cite{yeh2024enhanced,tang2025joint, Yin2026Graph,liu2025beamforming} are  not scalable with respect to the number of reflecting elements. The authors in \cite{S2025Optimization} considered an RIS-assisted cell-free system under imperfect CSI assumption and formulated a max-min uplink rate maximization problem. The system is represented by a heterogeneous graph comprising access point nodes, RIS reflecting element nodes, and user nodes. They proposed an advanced HGNN to jointly solve the problem.
The authors in \cite{le2025graph} considered a simultaneous transmitting and reflecting (STAR)-RIS-assisted MISO system and formulated a sum-rate maximization problem. They   modeled the system using a bipartite graph and solved the problem with a beamforming HGNN  based on a heterogeneous graph message-passing protocol, with guaranteed permutation equivariance and scalability.

% For the optimization objective of maximizing the weighted sum rate in multi-RIS multi-user mmWave systems, a heterogeneous GNN is adopted to model the base station, RISs and users as heterogeneous nodes with differentiated optimization features embedded, and fuse dual loss functions for joint learning of continuous and discrete variables to iteratively optimize RIS association and beamforming design. 

%For the optimization objective of maximizing the sum rate in RIS-aided multi-user downlink wireless communication systems, a graph attention network with unsupervised learning is adopted to model users as graph nodes and the channel matrix product as graph edges, and a two-stage network architecture combining GNN and MLP is designed to jointly optimize the transmit precoding matrix of the base station and the phase shift matrix of the RIS for efficient precoding design

\begin{table*}[t]
\centering
\caption{Comparison with existing works on GNN-enabled RIS-assisted systems.}\label{com}
\begin{tabular}{c||c|c|c|c|c|c|c|c|c}
%\hline
\hline
\multirow{2}{*}{\bf Ref.}  & \multirow{2}{*}{\bf Architecture} & \multirow{2}{*}{\bf Objective} &\multirow{2}{*}{\bf Multi-Stage}  &\multirow{2}{*}{\bf Heterogeneous}& \multirow{2}{*}{\bf Residual}& \multirow{2}{*}{\bf Attention} &  \multicolumn{3}{c}{{\bf Generalization}}    \\
\cline{8-10}
&&&&&&&LU&RIS&Eve\\
\hline\hline
\cite{Yang2024GATPrecoding}&GAT+MLP&SR&$\checkmark$&$\times$&$\times$&$\checkmark$&$\times$&$\times$&--\\
\cite{he2025ris}   & GAT+GCN &SR/EE &$\checkmark$ &$\times$ &$\checkmark$& $\checkmark$& $\checkmark$ &$\checkmark$&--\\
\cite{yeh2024enhanced}   & GCN+GAT &SR & $\times$ &$\checkmark$&$\times$&$\checkmark$&$\times$& $\times$&--\\

\cite{tang2025joint}  & HGNN &SR & $\times$ &$\checkmark$ &$\checkmark$&$\times$& $\checkmark$&$\times$&--\\

\cite{Yin2026Graph}&HGNN&EE&$\times$&$\checkmark$&$\checkmark$&$\times$&$\checkmark$&$\times$&--\\

\cite{liu2025beamforming}   & HGNN &WSR & $\times$ &$\checkmark$ &$\checkmark$& $\times$&$\times$&$\checkmark$ &--\\

\cite{S2025Optimization}&HGNN&Max-Min&$\times$&$\checkmark$&$\times$&$\times$&$\checkmark$&$\checkmark$&--\\

\cite{le2025graph} & HGNN & SR&$\times$ &$\checkmark$ &$\times$& $\times$&  $\checkmark$&$\checkmark$ &--\\

%\textcolor{blue}{\cite{liu2023cooperative}}&HGNN&WSR&$\times$&$\checkmark$&$\checkmark$&$\times$&$\checkmark$&$\checkmark$&--\\

\hline

\cite{Zhang2026Secrecy}&GNN&SSR&$\times$&$\checkmark$&$\checkmark$&$\times$&$\checkmark$&$\times$&$\times$\\
\cite{liang2025heterogeneous}   & CO-GNN &SSR &$\times$ &$\checkmark$ &$\times$ & $\checkmark$&  $\checkmark$&$\times$&$\checkmark$\\
\hline
{\bf Our work}  & HGNN & SSE
& $\checkmark$ &$\checkmark$ &$\checkmark$ & $\checkmark$&  $\checkmark$&$\checkmark$&$\checkmark$\\
\hline
\end{tabular}
\end{table*}

\subsection{GNN-Enabled RIS-Aided PLS}
% To make up for the shortcomings of convex optimization and DRL methods, data-driven learning optimization methods have become the mainstream research direction for RIS system optimization in recent years. This method combines the strong fitting ability of machine learning with the high solution accuracy of convex optimization, and realizes efficient optimization of RIS systems through the integration of data-driven prediction/approximation and exact solutions based on mathematical optimization. As a powerful data-driven learning tool, GNN has unique advantages in modeling the topological structure and interaction relationship of complex systems, and has been widely applied to various optimization tasks of wireless communication systems, such as resource allocation, beamforming, and channel estimation.  Numerous scholars have thus proposed RIS optimization methods based on heterogeneous graph neural networks (HGNNs), with representative research advances as follows, 

There are a limited number of works focusing on GNN-enabled RIS-aided PLS. 
 The authors in \cite{Zhang2026Secrecy} presented an RIS-assisted integrated sensing and communication (ISAC) system as a heterogeneous graph, which models LUs and an Eve as receiver nodes, and the RIS as an additional node. They utilized an HGNN to maximize the minimum secrecy rate. The authors in \cite{liang2025heterogeneous} considered an RIS-assisted non-orthogonal multiple access (NOMA) system encompassing downlink, uplink, and eavesdropping links. They modeled the system as a heterogeneous graph involving a BS node, a RIS node, and multiple LU nodes and Eve nodes. They proposed a combinatorial optimization GNN (CO-GNN) to maximize sum secrecy rate. However, the models proposed in \cite{Zhang2026Secrecy} and \cite{liang2025heterogeneous} rely on a single-stage framework and lack scalability with respect to the number of reflecting elements. 

%proposes a Graph Neural Network GNN-based unsupervised optimization framework for RIS-assisted Integrated Sensing and Communication systems, addressing physical layer security enhancement and secrecy rate maximization by modeling legitimate users and the eavesdropper as receiver nodes and the RIS as an additional node to form an attributed graph, effectively aggregating inter-node interference information to adaptively optimize the transmit beamforming vector and RIS phase shift matrix and prevent signal interception by eavesdroppers.

From the related works, we derive the following observations for designing GNN-enabled RIS-assisted systems: 1) both the multi-stage framework and heterogeneous graph representation can enhance the model's expressive power; 2) sophisticated graph modeling enables strong scalability. Furthermore, the design of GNN architecture should be well aligned with the adopted framework and graph representation, which is crucial to the model performance. Motivated by the above insights, we develop a two-stage HGNN for efficient and scalable design in RIS-aided PLS systems. %\textcolor{blue}{Compared with existing methods, this HGNN considers splitting subgraph types based on meta-paths instead of letting the network distinguish the types by itself, and utilizes the graph attention mechanism while incorporating channel values into residuals for use in RIS}.

%Although notable advances have been made in GNN-based RIS optimization research, evident research gaps remain in existing studies on RIS-aided PLS systems. On the one hand, the majority of HGNN-based RIS optimization methods are tailored for single-functional service systems (e.g., pure information transmission or standalone energy harvesting), whereas research on HGNN modeling for RIS-assisted PLS systems with dual-functional services—namely providing information transmission for legitimate users and countering active jamming interference for eavesdroppers is still quite scarce. On the other hand, the existing two-stage deep learning framework and HGNN model are typically investigated in isolation, and few studies have integrated the respective advantages of the two-stage framework and HGNN—this constitutes the core research problem to be addressed in this paper.

\section{System Model and Problem Formulation}

\begin{figure}[t]
\begin{center}
\centerline{\includegraphics[ width=.48\textwidth]{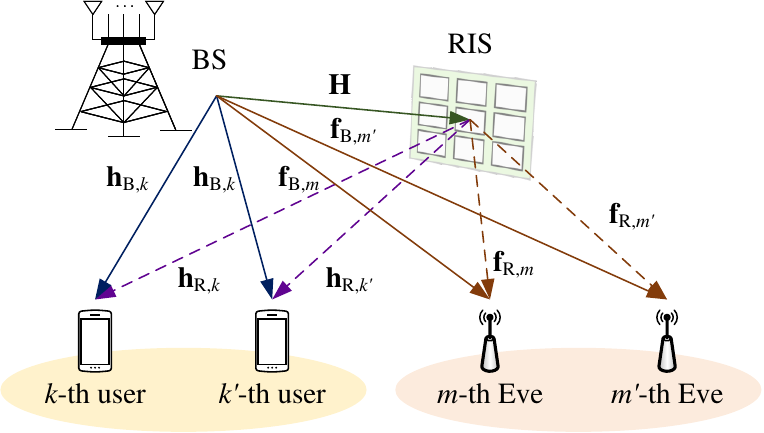}}
\caption{An RIS-aided MISO PLS system, where a BS serves multiple LUs in the presence of multiple Eves.}
\label{sys}
\end{center}
\end{figure}

\subsection{Signal Transmission}

Consider a passive RIS-assisted communication system as shown in Fig. \ref{sys} where an $N_{\rm T}$-antenna BS serves $K$ single-antenna LUs with the assistance of an RIS equipped with $L$  reflecting elements. $M$ single-antenna Eves intend to intercept information from the BS to LUs. For clarity, we use $k \in \mathcal{K} = \{1,2, ..., K\}$ to denote the $k$-th LU, $l \in \mathcal{L} = \{1,2, ..., L\}$ to denote the $l$-th reflecting element and $m \in \mathcal{M} = \{1,2, ..., M\}$ to denote the $m$-th Eve.

In each time slot, the transmit signal of the BS is given by
 \begin{flalign}\label{trans_signal}
{\bf x} = \sum\nolimits_{k\in{\cal K}}{\bf w}_k s_k + \sum\nolimits_{m\in{\cal M}}{\bf z}_m s_0,
\end{flalign}
where $s_k$ and  ${\bf w}_k \in \mathbb{C}^{N_{\rm T}} $ denote the information-bearing signal and the beamforming vector for the $k$-th LU, respectively, and  $s_0\in \mathbb{C}$ and ${\bf z}_m \in \mathbb{C}^{N_{\rm T}}$  denote the AN signal and the AN vector for the $m$-th Eve, respectively. Without loss of generality, it is assumed that ${{\mathbb E}}\{ {{{| {s_k} |}^2}} \} = 1$ and ${{\mathbb E}}\{ {{{| {s_0} |}^2}} \} = 1$.

The received signal at the $k$-th LU is given by
\begin{flalign}\label{rec_sig}
{y_k}= \underbrace {\left( {{\bf{h}}_{{\rm B},k}^H{\rm{ + }}{\bf{h}}_{{\rm R},k}^H{\bm\Phi} {{\bf{H}}}} \right)}_{\triangleq\widetilde{{\bf h}}^H_k\left({\bm\Phi}\right)\in{\mathbb C}^{N_{\rm T}}}\left(\sum\limits_{k^\prime\in \mathcal{K}}{\bf w}_{k^\prime} s_{k^\prime} + \sum\limits_{m^\prime\in \mathcal{M}}{\bf z}_{m^\prime} s_0\right) + {n_k},
\end{flalign}
where ${\bf h}_{{\rm B},k}\in{\mathbb C}^{N_{\rm T}}$, ${\bf h}_{{\rm R},k}\in{\mathbb C}^{L}$, and ${\bf H}\in{\mathbb C}^{L\times N_{\rm T}}$ respectively denote the direct CSI between the BS and the $k$-th LU, the CSI between the RIS and the $k$-th LU and the CSI between the BS and the RIS. ${\bm\Phi}= {\rm diag}\{[e^{j\phi_{1}},e^{j\phi_{2}},...,e^{j\phi_{L}}]\} $ denotes the phase shift matrix, where $\phi_{l} \in [0,2\pi)$ denotes the phase shift of the $l$-th reflecting element. $n_k\sim\mathcal{CN}( {0,{\sigma_k ^2}})$ denotes the additive white Gaussian noise (AWGN) at the $k$-th LU. The received information rate at the $k$-th LU is given by \eqref{r_u}.

\begin{figure*}
\begin{flalign}\label{r_u}
R_k\left( {\left\{ {{{\bf{w}}_{k{^\prime}}}}, {{{\bf{z}}_{m{^\prime}}}} \right\},{\bm\Phi}} \right) = {\log _2}\left( {1 + \frac{{{{\left| {{\widetilde{\bf h}^H_k\left({\bm\Phi}\right)}{{\bf{w}}_k}} \right|}^2}}}{{\sum\nolimits_{{{k^\prime}\in{\cal K}}/\{k\}} {{{\left| {{\widetilde{\bf h}^H_k\left({\bm\Phi}\right)}{{\bf{w}}_{k^\prime}}} \right|}^2}}+ \sum\nolimits_{{m^\prime} \in {\cal M}} {{{\left| {{\widetilde{\bf{h}}_k^H }\left({\bm\Phi}\right){{\bf{z}}_{m^\prime}}} \right|}^2}} + \sigma _k^2}}} \right)
\end{flalign}
\hrule
\end{figure*}

Similarly, the received signal at the $m$-th Eve is given by
\begin{flalign}
&{y_{{\rm E},m}}=\\
&\underbrace {\left( {{\bf{f}}_{{\rm B},m}^H{\rm{ + }}{\bf{f}}_{{\rm R},m}^H{\bm\Phi} {{\bf{H}}}} \right)}_{\triangleq\widetilde{\bf f}^H_m\left({\bm\Phi}\right)\in{\mathbb C}^{N_{\rm T}}}\left(\sum\limits_{k^\prime\in \mathcal{K}}{\bf w}_{k^\prime} s_{k^\prime} + \sum\limits_{m^\prime\in \mathcal{M}}{\bf z}_{m^\prime} s_0\right) + {n_{{\rm E},m}},\nonumber
\end{flalign}
where ${\bf f}_{{\rm B},m}\in{\mathbb C}^{N_{\rm T}}$ and ${\bf f}_{{\rm R},m}\in{\mathbb C}^{L}$  respectively denote the direct CSI from the BS and the RIS to the $m$-th Eve, and ${n_{{\rm E},m}}\sim\mathcal{CN}( {0,{\sigma_{{\rm E},m} ^2}})$ denotes the AWGN at the $m$-th Eve. If the $m$-th Eve intends to intercept the $k$-th LU, the information leakage data rate is expressed as \eqref{r_eve}.

\begin{figure*}
\begin{flalign}\label{r_eve}
R_{{\rm E},m,k}\left( {\left\{ {{{\bf{w}}_{k{^\prime}}}}, {{{\bf{z}}_{m{^\prime}}}} \right\},{\bm\Phi}} \right) = {\log _2}\left( {1 + \frac{{{{\left| {\widetilde{\bf f}^H_m\left({\bm\Phi}\right){{\bf{w}}_k}} \right|}^2}}}{{\sum\nolimits_{k^{\prime} \in {\cal K}/\{k\}} {{{\left| {\widetilde{\bf f}^H_m\left({\bm\Phi}\right){{\bf{w}}_{k^{\prime}}}} \right|}^2}}+ \sum\nolimits_{m^{\prime} \in {\cal M}} {{{\left| {\widetilde{\bf f}^H_m\left({\bm\Phi}\right){{\bf{z}}_{m^{\prime}}}} \right|}^2}} + \sigma _{{\rm E},m}^2}}} \right)
\end{flalign}
\hrule
\end{figure*}

The secrecy rate of the $k$-th LU is expressed as
\begin{flalign}
&R_{{\rm Sec},k}\left( {\left\{ {{{\bf{w}}_{k{^\prime}}}}, {{{\bf{z}}_{l{^\prime}}}} \right\},{\bm\Phi}} \right) =\\
& {\left[ {{R_k}\left( {\left\{ {{{\bf{w}}_{k{^\prime}}}}, {{{\bf{z}}_{m{^\prime}}}} \right\},{\bm\Phi}} \right) - \mathop {\max }\limits_{m} \left\{ {{R_{{\rm{E}},{m},k}}\left( {\left\{ {{{\bf{w}}_{k^\prime}}}, {{{\bf{z}}_{m^\prime}}} \right\},{\bm\Phi}} \right)} \right\}} \right]^ + }\nonumber.
\end{flalign}

\subsection{Problem Formulation}

Our goal is to jointly optimize ${\{ {{{\bf{w}}_{k^\prime}}},{\bf z}_{m^\prime},{{{\bm{\Phi}}}} \}}$ to maximize the system SEE under the constraint of the power budget denoted by $P_{\rm max}$. Mathematically, this problem is formulated as
\begin{subequations}\label{p0}
\begin{align}
&\mathop {\max }\limits_{{\left\{ {{{\bf{w}}_{k{^\prime}},{\bf z}_{m{^\prime}}}} \right\},{\bm\Phi}}} \frac{\sum\nolimits_{k \in {\cal K}}  R_{{\rm Sec},k}\left( {\left\{ {{{\bf{w}}_{k{^\prime}}}}, {{{\bf{z}}_{m{^\prime}}}} \right\},{\bm\Phi}} \right)}{\sum\nolimits_{k \in {\cal K}} {{{\left\| {{{\bf{w}}_{k}}} \right\|}^2}}  + \sum\nolimits_{m \in {\cal M}} {{{\left\| {{{\bf{z}}_{m}}} \right\|}^2}}+P_{\rm C}} \label{p0:a} \\
{\rm s.t.}~&\sum\nolimits_{k \in {\cal K}} {{{\left\| {{{\bf{w}}_k}} \right\|}^2}} + \sum\nolimits_{m \in {\cal M}} {{{\left\| {{{\bf{z}}_m}} \right\|}^2}}  \le {P_{\rm max}}, 
\label{p0:b}\\
&\phi_l \in [0,2\pi), l \in {\cal L},\label{p0:c}
\end{align}
\end{subequations}
where $P_{\rm C}$ denotes the constant power consumption. 

To solve Problem \eqref{p0}, we leverage the learning-to-optimize approach and propose a two-stage HGNN to map CSI to $\{ {\bf{w}}_{k^\prime}, {\bf z}_{m^\prime}, {\bm{\Phi}} \}$.

\section{Graph Representation}

\newcommand{\RNum}{\mathbb{R}}
\newcommand{\CNum}{\mathbb{C}}
\newcommand{\Concat}{\operatorname{Concat}}
\newcommand{\sigmoid}{\operatorname{sigmoid}}
\newcommand{\expf}{\operatorname{exp}}
\newcommand{\padding}{\operatorname{padding}}
\newcommand{\card}[1]{\left\vert #1 \right\vert}
\newcommand{\Nbr}[1]{\mathcal{N}(#1)}

Graph representation is utilized to organize system parameters into graph-structured data and to define readout operations. The proposed model adopts a two-stage framework, where the two stages employ different heterogeneous graph representation methods to represent the considered system. Notably, compared with existing graphs designed for  non-RIS and non-PLS systems \cite{li_twc_2024,he2025ICGNN}, the proposed graph captures the heterogeneity among different system entities, including  the RIS, LUs, and Eves. Compared with most existing graphs representing RIS-assisted PLS systems \cite{Zhang2026Secrecy,liang2025heterogeneous}, the proposed graph identifies distinct RIS reflecting elements rather than regarding all elements as a single node. 

Mathematically, we define a heterogeneous graph as 
\begin{flalign}
    {\cal G} = \left\{{\cal V}_1,\ldots,{\cal V}_N, {\cal E}_1,\ldots,{\cal E}_E \right\},
\end{flalign}
where $N$ and $E$ denote the number of node types and edge types, respectively, and ${\cal V}_n$ ($n\in\{1,2,\ldots,N\}$) and ${\cal E}_e$ ($e\in\{1,2,\ldots,E\}$) represent the node set of the  $n$-th type and the edge set of the $e$-th type. Furthermore, the node feature matrix and edge feature matrix for all nodes in ${\cal V}_n$ and all edges in ${\cal E}_e$ are defined by ${\bf X}_n$ and ${\bf Y}_e$, respectively.

\subsection{Stage 1: Bipartite Graph Representation}

\begin{figure}[!t]
  \centering
  \includegraphics[width=\columnwidth]{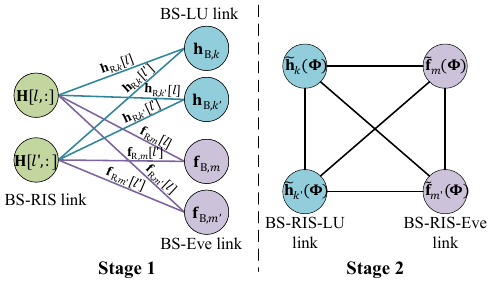} 
  \caption{Heterogeneous graph presentation and key node feature components: 1) Stage 1 constructs a bipartite graph consisting of three types of nodes corresponding to the BS-LU links, the BS-Eves links, and the BS-RIS links; 2) Stage 2 constructs a fully-connected graph consisting of two types of nodes corresponding to the BS-RIS-LU links and BS-RIS-Eve links.}
  \label{fig:stageExample}
\end{figure}

\begin{figure}[!t]
  \centering
  \includegraphics[width=\columnwidth]{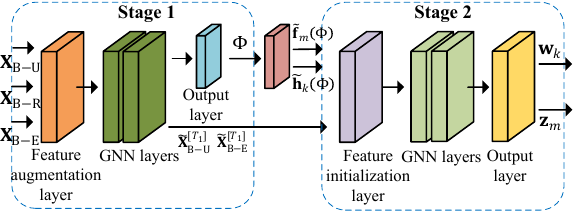} 
  \caption{Neural architecture of the two-stage HGNN: 1) Stage 1 comprises a feature augmentation layer, $T_1$ GNN layers, and an output layer. It maps {$\{{{\bf X}_{\rm B\text{-}R}, {\bf X}_{\rm B\text{-}U}, {\bf X}_{\rm B\text{-}E}}\}$} to {$\{\bm \Phi, {{\widetilde{\bf {\bf X}}^{[T_1]}_{\rm B\text{-}U}}}, {{\widetilde{\bf {\bf X}}^{[T_1]}_{\rm B\text{-}E}}}\}$}. 2) Stage 2 consists of a feature initialization layer, $T_2$ GNN layers, and an output layer. It takes {$\{\widetilde{\bf h}_k({\bm\Phi}), \widetilde{\bf f}_m({\bm\Phi})\}$} and  {$\{{{\widetilde{\bf {\bf X}}^{[T_1]}_{\rm B\text{-}U}}}, {{\widetilde{\bf {\bf X}}^{[T_1]}_{\rm B\text{-}E}}}\}$} as inputs and outputs {$\{{{\bf w}}_{k}, {{\bf z}}_{m}\}$}.}
  \label{fig:overallProcess}
\end{figure}

\begin{figure*}[!t]
  \centering
  \includegraphics[width=\textwidth]{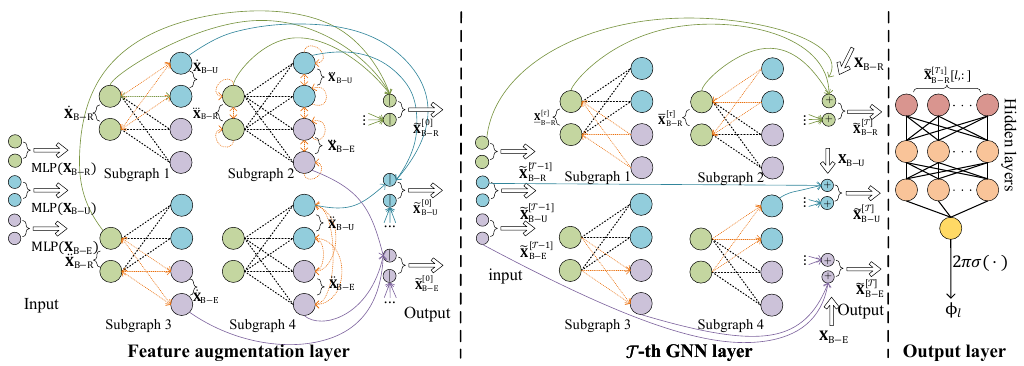} 
  \caption{Illustration of key processes in Stage 1: 1) The feature augmentation layer maps $\{{\bf X}_{\rm B\text{-}R}, {\bf X}_{\rm B\text{-}U},{\bf X}_{\rm B\text{-}E}\}$  to $\{\widetilde{\bf X}^{[0]}_{\rm B\text{-}R},\widetilde{\bf X}^{[0]}_{\rm B\text{-}U},\widetilde{\bf X}^{[0]}_{\rm B\text{-}E}\}$ via GAT on four subgraphs. 2) The $\tau$-th GNN layer maps  $\{\widetilde{\bf X}^{[\tau-1]}_{\rm B\text{-}R},\widetilde{\bf X}^{[\tau-1]}_{\rm B\text{-}U},\widetilde{\bf X}^{[\tau-1]}_{\rm B\text{-}E}\}$ combined with the residual terms $\{\widetilde{\bf X}^{[\tau-1]}_{\rm B\text{-}R},\widetilde{\bf X}^{[\tau-1]}_{\rm B\text{-}U},\widetilde{\bf X}^{[\tau-1]}_{\rm B\text{-}E},{\bf X}_{\rm B\text{-}R}, {\bf X}_{\rm B\text{-}U},{\bf X}_{\rm B\text{-}E}\}$ to $\{\widetilde{\bf X}^{[\tau]}_{\rm B\text{-}R},\widetilde{\bf X}^{[\tau]}_{\rm B\text{-}U},\widetilde{\bf X}^{[\tau]}_{\rm B\text{-}E}\}$ via GAT on four subgraphs. 3) The output layer maps $\{{{\widetilde{ {\bf X}}^{[T_1]}_{\rm B\text{-}R}}}[l,:]\}$ to $\{\phi_{l}\}$ using an MLP.}
  \label{fig:stage1Process}
\end{figure*}

\begin{figure*}[!t]
  \centering
  \includegraphics[width=\textwidth]{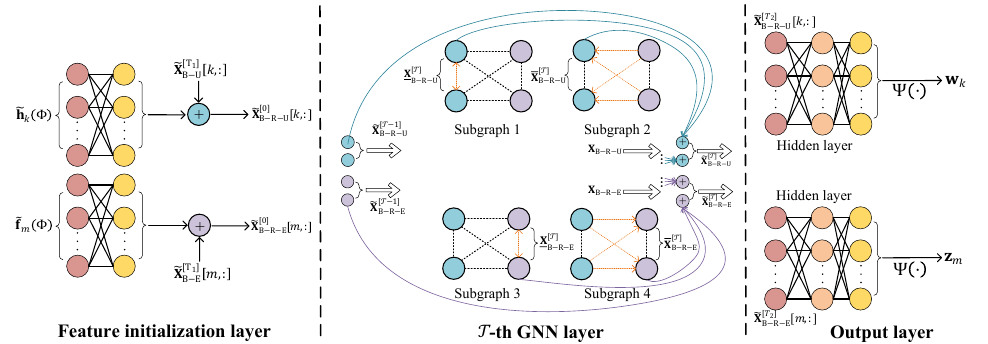} 
  \caption{Illustration of key processes in Stage 2: 1) The feature initialization layer employs two separate MLPs to map $\{{\widetilde{\bf h}_k},{\widetilde{\bf f}_m}\}$ to intermediate values, respectively. These values are then summed with $\{{{\widetilde{{\bf X}}^{[T_1]}_{\rm B\text{-}U}}}[k,:],{{\widetilde{{\bf X}}^{[T_1]}_{\rm B\text{-}E}}}[m,:]\}$ to obtain $\{\widetilde{\mathbf{X}}^{[0]}_{\rm B\text{-}R\text{-}U}[k,:],\widetilde{\mathbf{X}}^{[0]}_{\rm B\text{-}R\text{-}E}[m,:]\}$. 2)
The $\tau$-th GNN layer maps $\{\widetilde{\mathbf{X}}^{[\tau-1]}_{\rm B\text{-}R\text{-}U},\widetilde{\mathbf{X}}^{[\tau-1]}_{\rm B\text{-}R\text{-}E}\}$ combined with the residual terms $\{\widetilde{\mathbf{X}}^{[\tau-1]}_{\rm B\text{-}R\text{-}U},\widetilde{\mathbf{X}}^{[\tau-1]}_{\rm B\text{-}R\text{-}E},{\bf X}_{\rm B\text{-}R\text{-}U},{\bf X}_{\rm B\text{-}R\text{-}E}\}$ to  $\{\widetilde{\mathbf{X}}^{[\tau]}_{\rm B\text{-}R\text{-}U},\widetilde{\mathbf{X}}^{[\tau]}_{\rm B\text{-}R\text{-}E}\}$ via GAT on four subgraphs. 
3) The output layer uses two separate MLPs to map  $\{\widetilde{\mathbf{X}}^{[T_2]}_{\rm B\text{-}R\text{-}U}[k,:],\widetilde{\mathbf{X}}^{[T_2]}_{\rm B\text{-}R\text{-}E}[m,:]\}$  to $\{{{\bf w}}_{k},{{\bf z}}_{m}\}$, respectively.}
  \label{fig:stage2Process}
\end{figure*}

The graph in Stage 1 is defined as 
\begin{flalign}
{{\cal G}^{(1)}}=\left\{{\cal V}^{(1)}_1,{\cal V}^{(1)}_2,{\cal V}^{(1)}_3,{\cal E}^{(1)}_1,{\cal E}^{(1)}_2\right\},
\end{flalign}
which is a bipartite and undirected graph with $3$ node types and $2$ edge types. The structure of {${\cal G}^{(1)}$} is illustrated in Fig. \ref{fig:stageExample}. The node definition, edge definition, and readout operation are detailed as follows.
\begin{itemize}
    \item \emph{Node Definition}: In ${\cal G}^{(1)}$, there are $3$ node types: 1) $|{\cal V}^{(1)}_1|=L$ BS-RIS link nodes, with the $l$-th node feature being ${\bf H}[l,:]\in {\mathbb C}^{N_T}$; 2) $|{\cal V}^{(1)}_2|=K$ BS-LU link nodes, with the $k$-th node feature being ${\bf h}_{{\rm B},k}$; 3) $|{\cal V}^{(1)}_3|=M$ BS-Eve link nodes, with the $m$-th node  feature  being ${\bf f}_{{\rm B},m}$. In summary, the node feature matrices corresponding to the three node types are defined as follows, respectively:
    \begin{flalign}
        &{\bf X}_{\rm B\text{-}R} = {\bf H}'\in{\mathbb R}^{M\times 2N_{\rm T}},\\
        &{\bf X}_{\rm B\text{-}U} = \left[ {\bf h}_{{\rm B},1}',{\bf h}_{{\rm B},2}';\ldots;{\bf h}_{{\rm B},K}' \right]^{\rm T}\in{\mathbb R}^{K\times 2N_{\rm T}}, \\
        &{\bf X}_{\rm B\text{-}E} =  \left[ {\bf f}_{{\rm B},1}',{\bf f}_{{\rm B},2}';\ldots;{\bf f}_{{\rm B},M}' \right]^{\rm T}\in{\mathbb R}^{L \times 2N_{\rm T}},
    \end{flalign}
    where 
    \begin{flalign}
        &{\bf H}' = {\rm Concat}\left({\rm real \left({\bf H}\right)},{\rm imag \left({\bf H}\right)}\right)\in{\mathbb R}^{M\times 2N_{\rm T}}, \\
        &{\bf h}_{{\rm B},i}' = {\rm Concat}\left({\rm real \left({\bf h}_{{\rm B},i}\right)},{\rm imag \left({\bf h}_{{\rm B},i}\right)}\right)\in{\mathbb R}^{2N_{\rm T}}, \\
        &{\bf f}_{{\rm B},i}' = {\rm Concat}\left({\rm real \left({\bf f}_{{\rm B},i}\right)},{\rm imag \left({\bf f}_{{\rm B},i}\right)}\right)\in{\mathbb R}^{2N_{\rm T}}.
    \end{flalign}
    \item \emph{Edge Definition}: In ${\cal G}^{(1)}$, there are $2$ undirected edge types:  1) $|{\cal  E}_1^{(1)}|=L\times K$ edges, each connecting one BS-RIS link node and one BS-LU link node, with the $\langle l,k \rangle$-th edge feature being ${\bf h}_{{\rm R},k}[l]$; 2) $|{\cal  E}_2^{(1)}|=L\times M$ edges, each connecting one BS-RIS link node and one BS-Eve link node, with the $\langle l,m \rangle$-th edge feature being ${\bf f}_{{\rm R},m}[l]$.
    \item \emph{Readout Operation}: The HGNN in Stage 1 updates the features of $(L+K+M)$ nodes. The updated feature  of $L$ BS-RIS link nodes are readout as ${\bm\Phi}$, which is fed into both Stage 2 (cf. \eqref{stagenodefea1}) and the  loss function (cf. \eqref{loss}). The updated features of each BS-LU link node and each BS-Eve link node are used  as augmented features (cf. \eqref{stagenodefea1} and \eqref{stagenodefea2}) for beamforming learning and AN learning, respectively, in Stage 2.

    %It includes $R$ RIS reflecting element-base station link nodes, $K$ legitimate user-base station link nodes, and $L$ eavesdropper-base station link nodes, with the total number of nodes satisfying $\card{\mathcal{V}}=R+K+L$.
    %\%item \textbf{Node features}: The feature of an RIS element link node is the channel vector $\bm{\mathcal{H}}_{i,:} \in \CNum^{N_T}$ between the base station and the $i$-th RIS element; the feature of a legitimate user link node is the channel vector $\mathbf{h}_{\rm B,k} \in \CNum^{N_T}$ between the base station and the $k$-th legitimate user; the feature of an eavesdropper link node is the channel vector $\mathbf{f}_{\rm B,l} \in \CNum^{N_T}$ between the base station and the $l$-th eavesdropper.
   % \item \textbf{Edge composition}: An undirected edge is constructed between each legitimate user/eavesdropper node and all RIS element nodes, with no connecting edges between nodes of the same type.
    %\item \textbf{Edge features}: The edge feature between an RIS element and a legitimate user is the channel value $h_{R,k}^{(r)} \in \CNum^{1}$, and the edge feature between an RIS element and an eavesdropper is the channel value $h_{R,l}^{(r)} \in \RNum^{1}$.
\end{itemize}

\subsection{Stage 2: Fully-Connected Graph Representation}

The graph in Stage 2 is defined as 
\begin{flalign}
{{\cal G}^{(2)}}=\left\{{\cal V}^{(2)}_1,{\cal V}^{(2)}_2,{\cal E}^{(2)}_1,{\cal E}^{(2)}_2,{\cal E}^{(2)}_3\right\},
\end{flalign}
which is a fully-connected and undirected graph. The structure of {${\cal G}^{(2)}$} is illustrated in Fig. \ref{fig:stageExample}. The node definition, edge definition, and readout operation are detailed as follows.
\begin{itemize}
    \item \emph{Node Definition}: In ${\cal G}^{(2)}$, there are $2$ node types: 1) $|{\cal  V}_1^{(2)}|=K$ BS-RIS-LU link nodes, where the $k$-th node feature (cf. \eqref{stagenodefea1}) consists of ${\widetilde{{\bf h}}_k({\bm\Phi})}$ and the augmented feature of the $k$-th BS-LU node yielded by Stage 1; 2) $|{\cal  V}_2^{(2)}|=M$ BS-RIS-Eve link nodes, where the $m$-th node feature consists of ${\widetilde{{\bf f}}_m({\bm\Phi})}$ and the augmented feature (cf. \eqref{stagenodefea2}) of the $m$-th BS-Eve node yielded by Stage 1. Here,  ${\bm\Phi}$ is  derived from Stage 1. 
    \item \emph{Edge Definition}: In ${\cal G}^{(2)}$, there are three types of edges which connect two BS-RIS-LU link nodes, one BS-RIS-LU link node and one BS-RIS-Eve link node, and two BS-RIS-Eve link nodes, respectively. But all edges are non-feature. 
    \item \emph{Readout Operation}: The HGNN in Stage 2 updates the features of $(K+M)$ nodes. The updated features of the $k$-th BS-RIS-LU link node and the $m$-th BS-RIS-Eve link node are used to readout relevant parameters for the $k$-th beamforming vector and $m$-th AN vector, respectively.  \end{itemize}

    \section{Two-Stage HGNN}

The overall computational process of the two-stage HGNN is illustrated in Fig. \ref{fig:overallProcess}, and the models of the two stages incorporate some similar operations based on the meta-paths  \cite{Wang2019Heterogeneous}. For clarity, we first define the common operations and then introduce the architectures of the models in the two stages, respectively.

\subsection{Common Operations}

%and one type of node type feature extraction operation are defined. 

\subsubsection{Subgraph Extraction Operations}

To flexibly extract the intra-graph features, two types of subgraph extraction operations are defined over a given graph $\cal G$. 

\begin{itemize}
    \item Unidirectional bipartite graph extraction operation: 
\begin{flalign}
    {\widetilde{\cal G}}_{\rm D-BG}= \psi_{\rm Uni}\left({\cal V}_n,{\cal V}_{n^{\prime}}\right),
\end{flalign}
where ${\cal V}_n$ and ${\cal V}_{n^{\prime}}$ denote two node sets of distinct types, and $ \widetilde{\cal G}_{\rm D-BG}$ is a subgraph of $\cal G$ that forms a directed bipartite graph. The extracted subgraph ${\widetilde{\cal G}}_{\rm D-BG}$ consists of two node sets, i.e., ${\cal V}_n$ and ${\cal V}_{n^{\prime}}$, along with directed\footnote{Even for an undirected $\cal G$, a directed subgraph can be constructed, since an undirected edge can be treated as bidirectional edges sharing the identical features.} edges from each node in ${\cal V}_n$ to each node in ${\cal V}_{n^{\prime}}$. The nodes and edges in ${\widetilde{\cal G}}_{\rm D-BG}$ have consistent features with those in $\cal G$.

 \item Bidirectional bipartite graph extraction operation:
 \begin{flalign}
    {\widetilde{\cal G}}_{\rm UD-BG}= \psi_{\rm Bi}\left({\cal V}_n,{\cal V}_{n^{\prime}}\right),
\end{flalign}
where ${\widetilde{\cal G}}_{\rm UD-BG}$ is a subgraph of $\cal G$ that forms an undirected bipartite graph. The only difference between ${\widetilde{\cal G}}_{\rm UD-BG}$ and ${\widetilde{\cal G}}_{\rm D-BG}$ is that the edges in ${\widetilde{\cal G}}_{\rm UD-BG}$ are  undirected.

\end{itemize}
%The parameters $A, B \subseteq \{ \text{RIS}, \text{USER}, \text{EVE} \}$ are non-empty at the same time, representing the source/target node type sets respectively.

%Define $\cal A$ and $\cal B$ by two different sets of one  type of node. %\textcolor{red}{which cannot be non-empty at the same time}.

\subsubsection{Fully-Connected Graph Construction Operation} To exploit the similarity among nodes with the same type, the fully-connected graph construction operation is defined by
\begin{flalign}
    {\widetilde{\cal G}}_{\rm UD-FC}= \psi_{\rm FC}\left({\cal V}_n\right),
\end{flalign}
where ${\widetilde{\cal G}}_{\rm UD-FC}$ is a fully-connected and undirected graph consisting of nodes in ${\cal V}_n$ with undirected and feature-free edges connecting these nodes. The nodes inherit the features from their original $\cal G$.

\subsubsection{Graph Attention Operators} To distinguish the interactions between nodes, two types of graph attention operators are defined, which adapt to different graph-structured features. The output dimensions of the both operators are uniformly set to $D \times B$, where $D$ and $B$ denote the number of attention heads and the projection dimension of a single head, respectively.
\begin{itemize}
    \item Edge-based graph attention operator:
    \begin{flalign}
        \left\{{\bf X}_n\right\}_{n\in\{1,2,\ldots,N^{\prime}\}} := {\upsilon}_{\rm EAtt}\left({\cal G},\left\{{\bf X}_n \right\}\right),
    \end{flalign}
    where $N^{\prime}\le N$ denotes the number of target node types. 
    \item Edge-free graph attention operator:
        \begin{flalign}
        \left\{{\bf X}_n\right\}_{n\in\{1,2,\ldots,N^{\prime}\}}:= {\upsilon}_{\rm Att}\left({\cal G},\left\{{\bf X}_n \right\}\right).
    \end{flalign}

\end{itemize}

The detailed processes of the two attention operators are detailed in the Appendix A and B. Notably, the learnable parameters and output dimensions of both operators are associated with the input graph.

\subsection{Stage 1: Phase Shift Matrix Learning over ${\cal G}^{(1)}$}

In Stage 1, we utilize an HGNN to learn the phase shift matrix over ${\cal G}^{(1)}$ through three sequential modules: 1) the feature augmentation layer; 2) $T_1$ GNN layers; and 3) the phase shift output layer. The detailed computational process of the HGNN is illustrated in Fig. \ref{fig:stage1Process}.

\subsubsection{Feature Augmentation Layer} This module is designed to enhance the initial input features, yielding the inputs for the first GNN layer, which are defined by $\widetilde{\bf X}^{[0]}_{\rm B\text{-}U}$, $\widetilde{\bf X}^{[0]}_{\rm B\text{-}E}$, and $\widetilde{\bf X}^{[0]}_{\rm B\text{-}R}$. 

We first apply a single-layer MLP to enhance the initial dimension, which expands the feature space and improves the model's expressive capacity before subsequent feature processing.
\begin{flalign}
&{\mathbf{X}}'_{\rm B\text{-}R}\left[l,:\right] = {\rm LeakyReLU}\left({\mathbf W}_{1}{\mathbf{X}}_{\rm B\text{-}R}\left[l,:\right]\right),\\
&{\mathbf{X}}'_{\rm B\text{-}U}\left[k,:\right] = {\rm LeakyReLU}\left({\mathbf W}_{1}{\mathbf{X}}_{\rm B\text{-}U}\left[k,:\right]\right),\\
&{\mathbf{X}}'_{\rm B\text{-}E}\left[m,:\right] = {\rm LeakyReLU}\left({\mathbf W}_{1}{\mathbf{X}}_{\rm B\text{-}E}\left[m,:\right]\right),
\end{flalign}
where ${\mathbf W}_{1}$ is a learnable weight matrix.

We employ graph attention operations to obtain
\begin{flalign}
&\left\{{{\dot{\bf {\bf X}}_{\rm B\text{-}R}}},{\dot{\bf {\bf X}}_{\rm  B\text{-}U}}\right\} ={\upsilon}_{\rm EAtt}\left(\psi_{\rm Bi}\left({\cal V}^{(1)}_1,{\cal V}_{2}^{(1)}\right),{\bf X}_{\rm B\text{-}R}',{\bf X}_{\rm B\text{-}U}'\right),\\
&\left\{{{\ddot{\bf {\bf X}}_{\rm B\text{-}R}}},{\dot{\bf {\bf X}}_{\rm  B\text{-}E}}\right\} = {\upsilon}_{\rm EAtt}\left(\psi_{\rm Bi}\left({\cal V}^{(1)}_1,{\cal V}_{3}^{(1)}\right),{\bf X}_{\rm B\text{-}R}',{\bf X}_{\rm B\text{-}E} \right)',\\
&\left\{{{\ddot{\bf {\bf X}}_{\rm B\text{-}U}}},{\ddot{\bf {\bf X}}_{\rm  B\text{-}E}}\right\} = {\upsilon}_{\rm Att}\left(\psi_{\rm FC}\left({\cal V}^{(1)}_2,{\cal V}_{3}^{(1)}\right),{\bf X}_{\rm B\text{-}U}',{\bf X}_{\rm B\text{-}E}' \right),\\
&\left\{{{\dddot{\bf {\bf X}}_{\rm B\text{-}R}}},{\dddot{\bf {\bf X}}_{\rm  B\text{-}U}},{\dddot{\bf {\bf X}}_{\rm  B\text{-}E}}\right\} ={\upsilon}_{\rm Att}\left(\psi_{\rm FC}\left({\cal V}^{(1)}_1 \right)\cup\psi_{\rm FC}\left({\cal V}^{(1)}_2 \right)\right.\nonumber\\
&~~~~\left.\cup\psi_{\rm FC}\left({\cal V}^{(1)}_3 \right),{\bf X}_{\rm B\text{-}R}',{\bf X}_{\rm B\text{-}U}',{\bf X}_{\rm B\text{-}E}'\right),
\end{flalign}
and then, construct
\begin{flalign}
&\widetilde{\bf X}^{[0]}_{\rm B\text{-}U} = \Concat\left(\dot{\bf X}_{\rm B\text{-}U},\ddot{\bf X}_{\rm B\text{-}U},\dddot{\bf X}_{\rm B\text{-}U} \right),\\
&\widetilde{\bf X}^{[0]}_{\rm B\text{-}E} = \Concat\left(\dot{\bf X}_{\rm B\text{-}E},\ddot{\bf X}_{\rm B\text{-}E},\dddot{\bf X}_{\rm B\text{-}E} \right),
\\&\widetilde{\bf X}^{[0]}_{\rm B\text{-}R} = \Concat\left(\dot{\bf X}_{\rm B\text{-}R},\ddot{\bf X}_{\rm B\text{-}R},\dddot{\bf X}_{\rm B\text{-}R} \right).
\end{flalign}

\subsubsection{GNN Layers} $T_1$ GNN layers update the node feature matrix for each node type using graph attention operators and residual connections.   

At the $\tau$-th layer, the node feature matrices are first computed as
\begin{flalign}
    \label{subgraph1eq}
    &{{\widetilde{\bf {\bf X}}^{[\tau]}_{\rm B\text{-}U}}} ={\upsilon}_{\rm EAtt}\left(\psi_{\rm Uni}\left({\cal V}^{(1)}_1,{\cal V}_{2}^{(1)}\right),{{\widetilde{\bf {\bf X}}^{[\tau-1]}_{\rm B\text{-}R}}},{{\widetilde{\bf {\bf X}}^{[\tau-1]}_{\rm B\text{-}U}}}\right),\\
    \label{subgraph2eq}
    & {{\widetilde{\bf {\bf X}}^{[\tau]}_{\rm B\text{-}E}}} ={\upsilon}_{\rm EAtt}\left(\psi_{\rm Uni}\left({\cal V}^{(1)}_1,{\cal V}_{3}^{(1)}\right),{{\widetilde{\bf {\bf X}}^{[\tau-1]}_{\rm B\text{-}R}}},{{\widetilde{\bf {\bf X}}^{[\tau-1]}_{\rm B\text{-}E}}}\right),\\
    &{{\widetilde{\bf {\bf X}}^{[\tau]}_{\rm B\text{-}R}}} = {{\overline{\bf {\bf X}}^{[\tau]}_{\rm B\text{-}R}}} +  {{\underline{\bf {\bf X}}^{[\tau]}_{\rm B\text{-}R}}},
\end{flalign}
where  
\begin{flalign}
    \label{subgraph3eq}
       & {{\overline{\bf {\bf X}}^{[\tau]}_{\rm B\text{-}R}}} ={\upsilon}_{\rm EAtt}\left(\psi_{\rm Uni}\left({\cal V}_{2}^{(1)},{\cal V}^{(1)}_1\right),{{\widetilde{\bf {\bf X}}^{[\tau-1]}_{\rm B\text{-}U}}},{{\widetilde{\bf {\bf X}}^{[\tau-1]}_{\rm B\text{-}R}}}\right),\\
        \label{subgraph4eq}
     & {{\underline{\bf {\bf X}}^{[\tau]}_{\rm B\text{-}R}}} ={\upsilon}_{\rm EAtt}\left(\psi_{\rm Uni}\left({\cal V}_{3}^{(1)},{\cal V}^{(1)}_1\right),{{\widetilde{\bf {\bf X}}^{[\tau-1]}_{\rm B\text{-}E}}},{{\widetilde{\bf {\bf X}}^{[\tau-1]}_{\rm B\text{-}R}}}\right).
\end{flalign}

Then, these node feature matrices undergo residual connections \cite{Zheng2022Residual} to mitigate the gradient vanishing problem caused by layer stacking:
\begin{flalign}
\label{padding1}
&{{\widetilde{\bf {\bf X}}^{[\tau]}_{\rm B\text{-}U}}} := {{\widetilde{\bf {\bf X}}^{[\tau]}_{\rm B\text{-}U}}} + P\left({\bf X}_{\rm B\text{-}U}\right) + P\left({{\widetilde{\bf {\bf X}}^{[\tau-1]}_{\rm B\text{-}U}}}\right),\\
\label{padding2}
&{{\widetilde{\bf {\bf X}}^{[\tau]}_{\rm B\text{-}E}}} := {{\widetilde{\bf {\bf X}}^{[\tau]}_{\rm B\text{-}E}}} + P\left({\bf X}_{\rm B\text{-}E}\right) + P\left({{\widetilde{\bf {\bf X}}^{[\tau-1]}_{\rm B\text{-}E}}}\right),\\
\label{padding3}
&{{\widetilde{\bf {\bf X}}^{[\tau]}_{\rm B\text{-}R}}} := {{\widetilde{\bf {\bf X}}^{[\tau]}_{\rm B\text{-}R}}} + P\left({\bf X}_{\rm B\text{-}R}\right) + P\left({{\widetilde{\bf {\bf X}}^{[\tau-1]}_{\rm B\text{-}R}}}\right),
\end{flalign}
where $P(\cdot)$ is the zero-padding operation, aiming to align the dimensions of the matrices in a lightweight manner.

\subsubsection{Phase Shift Output Layer}
For the $l$-th reflecting  element, its phase shift is obtained by 
\begin{flalign}
\phi_{l} = &2\pi\times\sigmoid\left(\mathbf{c}_1 { \rm LeakyReLU}\left(\mathbf{W}_2 \right.\right.\nonumber\\
&\left.\left.{ \rm LeakyReLU}\left(\mathbf{W}_3{{\widetilde{\bf {\bf X}}^{[T_1]}_{\rm B\text{-}R}}}[l,:]\right)\right)\right),
\end{flalign}
where $\mathbf{c}_1$ , $\mathbf{W}_2$, and $\mathbf{W}_3$ are learnable parameters.  Then, the phase shift matrix is output as
\begin{flalign}
    {\bm\Phi}= {\rm diag}\left\{\left[e^{j\phi_{1}},e^{j\phi_{2}},...,e^{j\phi_{L}}\right]\right\}.
\end{flalign}

%where $\mathbf{w}_2\in {\mathbb R}^{b_1}$ , $\mathbf{W}_2 \in {\mathbb R}^{b_1 \times b_2}$, and $\mathbf{W}_3  \in {\mathbb R}^{b_2 \times b_3}$ are learnable weight vectors, $b_1$ and $b_2$ denote the number of neurons in the hidden layers, and $b_3$ denotes the output dimension of the $T_1$-th GNN layer. Then, the phase shift matrix is output as

%Finally, the equivalent channels $h_k$ and $f_l$ are solved by Equation \ref{eq:equi_channel} (to be defined), which are used as the initial node feature input $\mathbf{X}_{init}$ of graph $G_2$.

\subsection{Stage 2: Beamforming and AN Learning over ${\cal G}^{(2)}$}

In Stage 2, we utilize an HGNN to learn the beamforming vectors and AN vectors over ${\cal G}^{(2)}$ through three sequential modules: 1) the feature initialization layer; 2) $T_2$ GNN layers; and 3) the beamforming and AN output layer. The detailed computational process
of the HGNN is illustrated in Fig. \ref{fig:stage2Process}.

\subsubsection{Feature Initialization Layer} With the obtained ${\bm\Phi}$, the effective CSI for the $k$-th LU and the $m$-th Eve are derived  as ${\widetilde{{\bf h}}_k({\bm\Phi})}$ and ${\widetilde{{\bf f}}_k({\bm\Phi})}$. The node feature matrices for the two node types in ${\cal G}^{(2)}$ are initialized by fusing the  effective CSI  and the outputs of Stage 1, which are given respectively  by 
\begin{flalign}\label{stagenodefea1}
    &\widetilde{\mathbf{X}}^{[0]}_{\rm B\text{-}R\text{-}U}\left[k,:\right] = {\rm LeakyReLU}\left({\mathbf W}_4{\widetilde{\bf h}_k}\left({\bm\Phi}\right)\right) + {{\widetilde{\bf {\bf X}}^{[T_1]}_{\rm B\text{-}U}}}\left[k,:\right],\\
    &\widetilde{\mathbf{X}}^{[0]}_{\rm B\text{-}R\text{-}E}\left[m,:\right] = {\rm LeakyReLU}\left({\mathbf W}_5{\widetilde{\bf f}_m}\left({\bm\Phi}\right)\right) + {{\widetilde{\bf {\bf X}}^{[T_1]}_{\rm B\text{-}E}}}\left[m,:\right],\label{stagenodefea2}
\end{flalign}
where ${\bf W}_4$ and ${\bf W}_5$ are learnable weight matrices to align the dimensions of the matrices. 

\subsubsection{$T_2$ GNN Layers} $T_2$ GNN layers update the node feature matrix for each node type using graph attention operators and residual connections.   

At the $\tau$-th layer, the node feature matrices are first computed as
\begin{flalign}
    &{{\widetilde{\bf {\bf X}}^{[\tau]}_{\rm B\text{-}R\text{-}U}}} = {{\overline{\bf {\bf X}}^{[\tau]}_{\rm B\text{-}R\text{-}U}}} + {{\underline{\bf {\bf X}}^{[\tau]}_{\rm \rm B\text{-}R\text{-}U}}},\\
    &{{\widetilde{\bf {\bf X}}^{[\tau]}_{\rm B\text{-}R\text{-}E}}} = {{\overline{\bf {\bf X}}^{[\tau]}_{\rm B\text{-}R\text{-}E}}} + {{\underline{\bf {\bf X}}^{[\tau]}_{\rm \rm B\text{-}R\text{-}E}}},
\end{flalign}
where
\begin{flalign}
    &{{\overline{\bf {\bf X}}^{[\tau]}_{\rm B\text{-}R\text{-}U}}} ={\upsilon}_{\rm Att}\left(\psi_{\rm Uni}\left({\cal V}^{(2)}_1,{\cal V}_{2}^{(2)}\right),{{\widetilde{\bf {\bf X}}^{[\tau-1]}_{\rm B\text{-}R\text{-}U}}},{{\widetilde{\bf {\bf X}}^{[\tau-1]}_{\rm B\text{-}R\text{-}E}}}\right),\\
    & {{\overline{\bf {\bf X}}^{[\tau]}_{\rm B\text{-}R\text{-}E}}} ={\upsilon}_{\rm Att}\left(\psi_{\rm Uni}\left({\cal V}^{(2)}_2,{\cal V}_{1}^{(2)}\right),{{\widetilde{\bf {\bf X}}^{[\tau-1]}_{\rm B\text{-}R\text{-}E}}},{{\widetilde{\bf {\bf X}}^{[\tau-1]}_{\rm R\text{-}R\text{-}U}}}\right),\\
    &{{\underline{\bf {\bf X}}^{[\tau]}_{\rm B\text{-}R\text{-}U}}} ={\upsilon}_{\rm Att}\left(\psi_{\rm FC}\left({\cal V}^{(2)}_1 \right),{{\widetilde{\bf {\bf X}}^{[\tau-1]}_{\rm B\text{-}R\text{-}U}}}\right),\\
    &{{\underline{\bf {\bf X}}^{[\tau]}_{\rm  B\text{-}U\text{-}E}}} ={\upsilon}_{\rm Att}\left(\psi_{\rm FC}\left({\cal V}^{(2)}_2 \right),{{\widetilde{\bf {\bf X}}^{[\tau-1]}_{\rm B\text{-}R\text{-}E}}}\right).
\end{flalign}

% To improve the network's ability to distinguish different edge features, a semantic attention mechanism is introduced, and the importance weight of each subgraph is calculated through a fully connected neural network. For $\mathbf{X}_{UE,g}'$, the semantic attention coefficient of the $g$-th subgraph is:
% \begin{align}\label{eq:semantic_attention}
% \gamma_{g}= \frac{\expf\left(\frac{1}{K+L}\sum_{i\in{\cal K}\cup\mathcal{L}}{\bf q}^{T}\tanh(\bm\Theta {\bf x}_{i,g}+\bm\xi)\right)}{\sum_{j\in\{1,2\}}\expf\left(\frac{1}{K+L}\sum_{i\in\mathcal{K}\cup\mathcal{L}}{\bf q}^{T}\tanh(\bm\Theta {\bf x}_{i,j}+\bm\xi)\right)},
% \end{align}
% where ${\bf q}$, $\bm\Theta$ and $\bm\xi$ are learnable weights. The features of this layer are fused based on the semantic attention coefficients to obtain the output of this layer:
% \begin{align}\label{eq:output_SAL}
% {\bf x}_{i} = {\rm LeakyReLU} \left(\sum_{g \in \{1,2\}}\gamma_{g}{\bf x}_{i,g} \right)\in \RNum^{b}.
% \end{align}

The residual connections are introduced to alleviate the gradient vanishing problem, and the final output of each layer is
\begin{flalign}
\label{padding4}
&{{\widetilde{\bf {\bf X}}^{[\tau]}_{\rm B\text{-}R\text{-}U}}} := {{\widetilde{\bf {\bf X}}^{[\tau]}_{\rm B\text{-}R\text{-}U}}} + P\left({\bf X}_{\rm B\text{-}R\text{-}U}^{}\right) + P\left({{\widetilde{\bf {\bf X}}^{[\tau-1]}_{\rm B\text{-}R\text{-}U}}}\right),\\
\label{padding5}
&{{\widetilde{\bf {\bf X}}^{[\tau]}_{\rm B\text{-}R\text{-}E}}} := {{\widetilde{\bf {\bf X}}^{[\tau]}_{\rm B\text{-}R\text{-}E}}} + P\left({\bf X}_{\rm B\text{-}R\text{-}E}^{}\right) + P\left({{\widetilde{\bf {\bf X}}^{[\tau-1]}_{\rm B\text{-}R\text{-}E}}}\right),
\end{flalign}
where
\begin{flalign}
    &{\bf X}_{\rm B\text{-}R\text{-}U} \triangleq \left[ {\widetilde{\bf h}_1}\left({\bm\Phi}\right),{\widetilde{\bf h}_2}\left({\bm\Phi}\right);\ldots;{\widetilde{\bf h}_K}\left({\bm\Phi}\right) \right], \\
&{\bf X}_{\rm B\text{-}R\text{-}E} \triangleq  \left[ {\widetilde{\bf f}_1}\left({\bm\Phi}\right),{\widetilde{\bf f}_2}\left({\bm\Phi}\right);\ldots;{\widetilde{\bf f}_M}\left({\bm\Phi}\right) \right].
\end{flalign}

\subsubsection{Beamforming and AN Output Layer} We consider two output methods, i.e.,  beam-direct and  model-based. 

The beam-direct output enables $(K+L)$ nodes to share two fully-connected layers and one activation function to  directly map the final node embeddings ${{\widetilde{\bf {\bf X}}^{[T_2]}_{\rm B\text{-}R\text{-}U}}}$ and ${{\widetilde{\bf {\bf X}}^{[T_2]}_{\rm B\text{-}R\text{-}E}}}$ to the unconstrained beamforming vector and AN vector:
\begin{flalign}\label{eq:FC1}
{\widetilde{\bf w}}_{k}=f_2\left({\bf W}_6 { \rm LeakyReLU}\left( {\bf W}_7{{\widetilde{\bf {\bf X}}^{[T_2]}_{\rm B\text{-}R\text{-}U}}}\left[k,:\right]+{\bm \eta}_1\right)+ {\bm \eta}_2 \right),
\\\label{eq:FC2}
{\widetilde{\bf z}}_{m}=f_2\left({\bf W}_8 { \rm LeakyReLU}\left( {\bf W}_9{{\widetilde{\bf {\bf X}}^{[T_2]}_{\rm B\text{-}R\text{-}E}}}\left[m,:\right]+{\bm \eta}_3\right)+ {\bm \eta}_4 \right),
\end{flalign}
where ${\bf W}_6$, ${\bf W}_7$, ${\bf W}_8 $, ${\bf W}_9$, ${\bm\eta}_1$, ${\bm\eta}_2$, ${\bm\eta}_3$, and ${\bm\eta}_4$ are learnable weights, and $f_2(\cdot):\RNum ^ {2N_{\rm T}}\to \CNum ^ {N_{\rm T}}$ is a dimension reshaping function. Then, $\left\{\widetilde{\bf w}_{k}\right\}$ and  $\left\{\widetilde{\bf z}_{m}\right\}$ undergo the following operation to meet the total power budget constraint: 
\begin{flalign}\label{af}
\Psi \left( {{\bf{v}}}_i \right) 
= \sqrt {\frac{{{P_{\rm max }}}} {{\rm max}\left({{P_{\rm max }}},\sum\nolimits_{i^{\prime} \in {\cal K}  \cup  {\cal M}} {{\left\| {{{\bf{v}}_{i^{\prime}}}} \right\|}^2}\right)}}{{\bf{v}}_i},
\end{flalign}
where ${\bf v}_n \in {\{ {{{\bf{w}}_k}},{{{\bf{z}}_m}} \}}_{k \in {\cal K},m \in {\cal M}}$.

The model-based output utilizes the hybrid zero forcing (ZF) and maximum ratio transmission (MRT) method \cite{10851843},  which defines $({\alpha_k,p_k})$ and  $({\beta_m,p_m})$ in the Appendix C, where  $\alpha_k$ and $\beta_m$ represent the direction-related parameters, and $p_k$ and $p_m$ represents the power-related parameters. The model-based output employs two fully-connected layers along with one activation function to yield $({\alpha_k,p_k})$ for the $k$-th LU and $({\beta_m,p_m})$ for the $m$-th Eve. Specifically, to meet  the constraints, $\alpha_k$ and $\beta_m$ are activated via the Sigmoid function, while  $p_k$ and $p_m$ undergo the ReLU function along with
\begin{flalign}\label{af2}
\widetilde{\Psi} \left( p_i \right) 
= \sqrt {\frac{{{P_{\rm max }}}} {{\rm max}\left({{P_{\rm max }}},\sum\nolimits_{i^{\prime} \in {\cal K}  \cup  {\cal M}} {p_{i^{\prime}}}\right)}p_i},
\end{flalign}
where ${p}_i \in {\{ {{p_k}},{{p_m}} \}}_{k \in {\cal K},m \in {\cal M}}$.

\subsection{Loss Function}

We adopt unsupervised learning to train the proposed model, and the loss function is given by
\begin{flalign}\label{loss}
{\mathcal L} = -\frac{\sum\nolimits_{k \in {\cal K}}  \widetilde{R}_{{\rm Sec},k}\left( {\left\{ {{{\bf{w}}_{k^\prime}}}, {{{\bf{z}}_{m^\prime}}} \right\},{\bm\Phi}} \right)}{\sum\nolimits_{k \in {\cal K}} {{{\left\| {{{\bf{w}}_k}} \right\|}^2}}  + \sum\nolimits_{m \in {\cal M}} {{{\left\| {{{\bf{z}}_m}} \right\|}^2}}+P_{\rm C}},
\end{flalign}
where $ \widetilde{R}_{{\rm Sec},k}( {\{ {{{\bf{w}}_{k^\prime}}}, {{{\bf{z}}_{m^\prime}}}\},{\bm\Phi}})$ is given by 
\begin{flalign}
&{\widetilde R}_{{\rm Sec},k}\left( {\left\{ {{{\bf{w}}_{k^\prime}}}, {{{\bf{z}}_{l^\prime}}} \right\},{\bm\Phi}} \right) =\\
& { {{R_k}\left( {\left\{ {{{\bf{w}}_{k^\prime}}}, {{{\bf{z}}_{m^\prime}}} \right\},{\bm\Phi}} \right) -  \gamma \mathop{\max }\limits_m \left\{ {{R_{{\rm{E}},m,k}}\left( {\left\{ {{{\bf{w}}_{k^\prime}}}, {{{\bf{z}}_{m^\prime}}} \right\},{\bm\Phi}} \right)} \right\}}  }\nonumber,
\end{flalign}
where $\gamma$ is a hyperparameter, which is usually set to  $0.1$.

\subsection{Scalability Analysis}

Notably, all learnable parameters in the proposed two-stage HGNN are independent of the number of reflecting elements $L$, LUs $K$, and Eves $M$.  This property stems from the permutation equivariance inherent in the graph representations and graph operations employed in the model.

\subsubsection{Permutation-Equivariant Graph Representation}
The graph in Stage 1 incorporates three types of nodes corresponding to reflecting elements, LUs, and Eves,  where both LU nodes and Eve nodes are connected to reflecting element nodes. 
The  graph in Stage 2 is fully-connected, which consists of two types of nodes corresponding to LUs and Eves, where the features of both node types involve the phase shift matrix. The graphs utilized in both stages satisfy permutation equivariance, i.e., permuting the input order of the same node type does not alter the model’s output or optimization results. 
Such a property renders the model inherently adaptable to dynamic numbers of reflecting elements, LUs, and Eves. In particular, this can be achieved simply by adding or removing corresponding nodes in the graph, while the core message-passing mechanism remains unchanged and no retraining is necessitated.

\subsubsection{Permutation-Equivariant Graph Operation}
The key operator employed in both stages relies on the multi-head attention mechanism from GAT. Such a mechanism is inherently permutation-equivariant. That is, arbitrary permutations of the input node set, including the rearrangement of node indices and their associated features, yield identical permutations of the output node representations, without altering the model’s core computational logic or the relative relationships among the learned node embeddings. When new nodes are added to the graph, the model automatically integrates them into the message-passing procedure by computing attention weights between the newly introduced nodes and existing ones according to their feature similarities, without any modification to the pre-trained model parameters. Similarly, when nodes are removed, the model simply excludes the corresponding nodes from the message aggregation and attention computation steps, while the underlying message-passing mechanism remains unchanged.

%\subsubsection{Scale-Independent Learnable Parameters}
%In this work, all learnable parameters in both the GAT and the output layer are independent of the number of RIS elements L, users K, and eavesdroppers M.Consequently, the proposed model does not require retraining when adding or removing nodes from the graph.This property arises from the permutation equivariance of the GAT architecture.

\section{Numerical Results}
This section evaluates the proposed model. The structures of the proposed model are displayed in Table \ref{table:gnn}.

\begin{table}[t]
\centering
\caption{The structure of the two-stage  HGNN.}
\begin{tabular}{c|cccccc}
\hline
Stage& \multicolumn{2}{c|}{Type}   & \multicolumn{1}{c}{IFs} & \multicolumn{1}{c}{HNs} & \multicolumn{1}{c}{OFs} & \multicolumn{1}{c}{AHs} \\ \hline\hline
\multicolumn{1}{c|}{\multirow{4}{*}{1}} & \multicolumn{2}{c|}{FIL}   &  2$N_{\rm T}$    &   640   &   32   &    5  \\ \cline{2-7} 
\multicolumn{1}{c|}{}& \multicolumn{1}{c|}{\multirow{2}{*}{GL}} & \multicolumn{1}{c|}{1} &  480    &   $\times$   &   64   &  10    \\
\multicolumn{1}{c|}{}& \multicolumn{1}{c|}{} & \multicolumn{1}{c|}{2} &      640&    $\times$  &   64   &   10   \\ \cline{2-7} 
\multicolumn{1}{c|}{}& \multicolumn{2}{c|}{OL}    &    640  &   320, 128   &    1  &   $\times$   \\ \hline\hline
\multicolumn{1}{c|}{\multirow{4}{*}{2}} & \multicolumn{2}{c|}{FAL}   &   2$N_{\rm T}$   &   $\times$   &  640    &   $\times$   \\ \cline{2-7} 
\multicolumn{1}{c|}{}& \multicolumn{1}{c|}{\multirow{2}{*}{GL}} & \multicolumn{1}{c|}{1} &   640   &   $\times$   &   128   &  10    \\
\multicolumn{1}{c|}{}& \multicolumn{1}{c|}{} & \multicolumn{1}{c|}{2} &      1280&   $\times$   &   256   &  10    \\ \cline{2-7} 
\multicolumn{1}{c|}{}& \multicolumn{2}{c|}{OL}    &    2560  &   640   &      2$N_{\rm T}$&   $\times$   \\ \hline
\end{tabular}
\label{table:gnn}
\begin{tablenotes}
    \footnotesize
    \item GL/FAL/FIL/OL: GNN layer/Feature augmentation layer/Feature initialization layer/Output layer.
    \item IFs/OFs/HNs/AHs: Input features/Output features/Number of neurons in hidden layer/Number of attention heads.
\end{tablenotes}
\end{table}
\begin{table}[t]
  \centering
  \belowrulesep=-0.5pt
\aboverulesep=-0.5pt
    \caption{Training and test set configurations.}
  \begin{tabular}{c|c}
    \toprule
    {\bf Training set} & {\bf Test set} \\
    $(L_{\rm Tr},K_{\rm Tr},M_{\rm Tr})$  & $(L_{\rm Te},K_{\rm Te},M_{\rm Te})$   \\
    \hline\hline
    \multirow{7}{*}{$(8,3,4)$} 
                      & $(8,3,4)$  \\
    &  $(8,2,4)$   \\
                  & $(8,4,4)$  \\
                & $(7,3,4)$ \\
               & $(9,3,4)$ \\
                 & $(8,3,3)$  \\
                   & $(8,3,5)$ \\
    \hline 
     \#Sample ~{$9.6 \times 10^5$}  & \#Sample ~{$1000$} per set   \\
    \bottomrule
    
  \end{tabular}

  \label{tab:dataset_config}
\end{table}

\subsection{Experimental Setting}

\subsubsection{Simulation Scenario}

The BS is deployed at $(0,0)$ m, while the RIS is placed at $(10,0)$ m close to the BS to mitigate cascaded fading. The LUs are randomly distributed within a circular area centered at 
$(50,50)$ m with a radius of $10$ m, and the Eves are randomly distributed within a circular area centered at $(25,25)$ m with a radius of 
$10$ m. Notably, Eves are closer to the BS than LUs. The transmit antennas at BS are set as $N_{\rm T}=8$. 

\subsubsection{Channel Model}

The path loss of direct link (including ${\bf h}_{{\rm B},k}$ and ${\bf f}_{{\rm B},m}$ ) is modeled as $\sqrt{\rho d^{-{\alpha}}}$ where $d$ and $\rho$  denote the corresponding  Euclidean distance and the path loss per meter distance, respectively, and $\alpha$ denotes 
the fading exponent. The small-scale fading is modeled as Rayleigh fading.

The RIS-related CSI contains one LoS path and a large number of NLoS paths. Specifically, the CSI between the RIS and a receiver (either an LU or an Eve), or between the BS and the RIS, is modeled using Rician fading, which is given by
 \begin{flalign}
{\bf h} &= \sqrt{\rho {d}^{-{\alpha}}}\left(\sqrt{\frac{\beta}{1+\beta}}{\bf h}^{\rm LoS}\left(\phi\right)+\sqrt{\frac{1}{\beta+1}}{{\bf h}^{\rm NL}}\right),\\
{\bf H} &= \sqrt{\rho {d}^{-{\alpha}}}\left(\sqrt{\frac{\beta}{1+\beta}}{\bf H}^{\rm LoS}\left(\phi\right)+\sqrt{\frac{1}{\beta+1}}{{\bf H}^{\rm NL}}\right),
 \end{flalign}
where $\beta$ denotes the Rician factor, and the entries of ${\bf h}^{\rm NL}/{{\bf H}^{\rm NL}}$  are independently drawn from the complex Gaussian distribution with zero mean and unit variance, i.e.,  ${\cal CN}(0,1)$. The LoS components are expressed by the responses of the ULA. For an $N$-element ULA, the response is given by
\begin{flalign}
{\bf a}_N\left(\phi\right)=\left[1,e^{-j\frac{2\pi}{\lambda}\Delta\sin\left(\phi\right)},...,e^{-j\frac{2\pi}{\lambda}(N-1)\Delta\sin\left(\phi\right)}\right]^T,
\end{flalign}
where $\Delta$ and $\lambda$ denote antenna separation and wavelength, respectively, and 
$\phi$ denotes the angle of departure (AoD) or angle of arrival (AoA) of a signal. Then, ${{\bf h}^{\rm LoS}}$ and ${{\bf H}^{\rm LoS}}$ are respectively given by
\begin{flalign}
{{\bf h}^{\rm LoS}}&={\bf a}_{L}\left(\phi_{\rm AoD}\right), \\
{{\bf H}^{\rm LoS}}&={\bf a}_{L}\left(\phi_{\rm AoA}\right){\bf a}_{N_{\rm T}}^H\left(\phi_{\rm AoD}\right).
\end{flalign}

All default parameters are summarized in Table \ref{table:simulation_settings}.

\begin{table}[t]
\centering
\belowrulesep=-0.5pt
\aboverulesep=-0.5pt
\caption{Simulation Parameters.}
\begin{tabular}{c|c }
 \toprule
{\bf Notation}               & {\bf Values}               \\ \hline
\hline
BS location & $(0,0)$\\ \hline
RIS location & $(10,0)$\\ \hline
LU location & center $(50,50)$ radius $10$\\ \hline
Eve location & center $(25,25)$ radius $10$\\ \hline
Number of antennas & $N_{\rm T} = 8$\\ \hline
Central frequency & $f_{\rm c}=1.8$ GHz \\ \hline
 Constant power & $P_{\rm C}=0.5$ W\\ \hline
Antenna separation & $\Delta={\lambda}/{2}$\\ \hline
Wavelength & $\lambda = {c}/{f_{\rm c}}$\\ \hline
Rician factor  & $\beta=3$ dB  \\\hline
Fading exponent    & $\alpha=2.8$   \\\hline
Power budget & $P_{\max} = $ 30 dBm  \\\hline
Path loss per
meter & $\rho=-20$ dB \\\hline
Noise power & $\sigma^2 = -80$ dBm \\
  \bottomrule
%AoD and AoA  & $\phi_{\rm AoA}, \phi_{\rm AoD}\sim {\cal U}(0,2\pi)$ \\\hline\hline
%    & $||{\bf h}||^2/\sigma^2$ within $\{0,10,100\}$ dB \\\hline
\end{tabular}
\label{table:simulation_settings}
\end{table}

\subsubsection{Dataset and Metric} We prepare one training set with $9.6\times10^5$ samples and seven test sets with $1000$ samples each. The training set adopts the setting of $(L_{\rm Tr},K_{\rm Tr},M_{\rm Tr})=(8,3,4)$, while the configurations of the test sets are provided in Table \ref{tab:dataset_config}. 

All samples in both the training set and test sets are generated independently according to the channel model. Moreover, for each test sample, we employ the BCD method to obtain its optimal SEE as the label. During the test, we use the ratio of the SEE yielded by the DL model to the SEE label as the evaluation metric.

\subsubsection{Implementation Details}The learning rate is initialized as \num{1e-4}. The adaptive moment estimation is adopted as the optimizer during the training phase. The batch size is set to $256$ for $100$ epochs of training, and the learnable weights with the best performance are used as the training result. Our implementation is developed using Python 3.11.5 with PyTorch 2.1.0 on a computer with Intel(R) Core(TM) Ultra 9 275HX CPU and one NVIDIA GeForce RTX 5070 Ti Laptop GPU (12 GB of memory).

\subsubsection{Baselines}
The baselines employed to evaluate the effectiveness of the proposed two-stage HGNN are listed as follows:
\begin{itemize}
    \item MLP: A basic feed-forward neural network with a three-branch output architecture.
    \item CO-GNN: A GNN model with one RIS node and multiple LU nodes, which adopts a custom message-passing mechanism integrating mean-pooling, max-pooling, and MLP-based node feature update, similar to \cite{liang2025heterogeneous}.
    \item BHGNN: A GNN model with multiple RIS nodes, multiple LU nodes, and edge feature integration, which adopts a custom message-passing mechanism integrating sum-pooling, MLP-based node feature update, and iterative optimization update, similar to \cite{le2025graph}.
    
\end{itemize}

\begin{table*}[t]
\belowrulesep=-0.5pt
\aboverulesep=-0.5pt
\centering
\caption{Performance comparison.}
\label{tab:beamforming}
\begin{tabular}{c|c|c|ccccc|c}
\toprule
\multirow{2}{*}{$N_{\rm T}$} & Training Setting & Test Setting  & \multicolumn{3}{c}{Baselines} & \multicolumn{2}{c|}{Two-stage HGNN} &\multirow{2}{*}{Note}\\
\cmidrule(lr){2-2}\cmidrule(lr){3-3}\cmidrule(lr){4-6} \cmidrule(lr){7-8}
& {$(L_{\rm Tr},K_{\rm Tr},M_{\rm Tr})$} & {$(L_{\rm Te},K_{\rm Te},M_{\rm Te})$} &  MLP & CO-GNN & BHGNN & Beam-direct & \multicolumn{1}{c|}{Model-based} \\
\hline
\hline
\multirow{7}{*}{$8$} & \multirow{7}{*}{$(8,3,4)$} & {$(8,3,4)$} & 23.57\% & 61.61\% & 34.87\% & 90.76\% & 96.60\% \\
\cline{3-9}
& & {$(8,2,4)$} & 28.31\% & 50.28\% & 49.59\% & 82.01\% & 96.80\% & Scale to \\
&& {$(8,4,4)$} & -- & 25.35\% & 2.01\% & 45.20\% & 96.99\% &  $L$ \\
\cline{3-9}
&& {$(7,3,4)$} & -- & -- & 23.98\% & 89.54\% & 96.46\% & Scale to \\
&& {$(9,3,4)$} & -- & -- & 25.26\% & 89.73\% & 96.43\% & $K$ \\
\cline{3-9}
&& {$(8,3,3)$} & 25.72\% & 63.03\% & 37.30\% & 91.56\% & 97.27\% & Scale to \\
&& {$(8,3,5)$} & -- & 61.06\% & 34.14\% & 79.39\% & 95.93\% & $M$ \\
\hline
\hline
\multicolumn{3}{c|}{Inference time}& $0.73$ ms & $0.99$ ms & $2.3$ ms & $11.53$ ms & $11.64$ ms & \thead{{\normalsize $82.74$ s by}\\{\normalsize  CVXopt}} \\
\bottomrule
\end{tabular}
\end{table*}

\subsection{Effectiveness} %模型对比&消融

Table~\ref{tab:beamforming} compares the proposed two-stage HGNN with the baselines. For identical training and test settings, i.e., $(L_{\rm Tr},K_{\rm Tr},M_{\rm Tr})=(L_{\rm Te},K_{\rm Te},M_{\rm Te})$,  the proposed two-stage HGNN achieves a higher SEE using both the beam-direct and model-based output methods, where the model-based one can attain near-optimal performance. However, the existing baselines fail to yield satisfactory performance. Especially, the MLP achieves a performance near $20\%$, highlighting the necessity of developing task-suitable neural architectures. The two-stage HGNN, COGNN and BHGNN are scalable with respect to the number of reflecting elements and Eves. As observed, only the two-stage HGNN with the model-based output method can maintain a good performance. In contrast, the two-stage HGNN with the beam-direct output method suffers from a notable performance degradation when encountering a larger $L$ or $M$. Furthermore, the baselines exhibit even poorer performance. The two-stage HGNN and BHGNN are scalable with respect to the number of LUs $K$. In this scenario, both the beam-direct and model-based output methods maintain limited performance loss, whereas BHGNN does not scale well. Moreover, the two-stage HGNN requires longer inference time for the transmit design than the baselines, since it involves more sophisticated designs in both framework and architecture. Nevertheless, it still maintains millisecond-level efficiency, which is considerably faster than the CVXopt-based approach. In summary, the two-stage HGNN yields near-optimal performance, with a notable performance improvement over the baselines. Moreover, the two-stage HGNN, especially with the model-based output method, scales well with respect to the number of reflecting elements, LUs, and Eves, demonstrating its adaptability to dynamic wireless environments.

\begin{figure}[!t]
  \centering
  \includegraphics[width=\columnwidth]{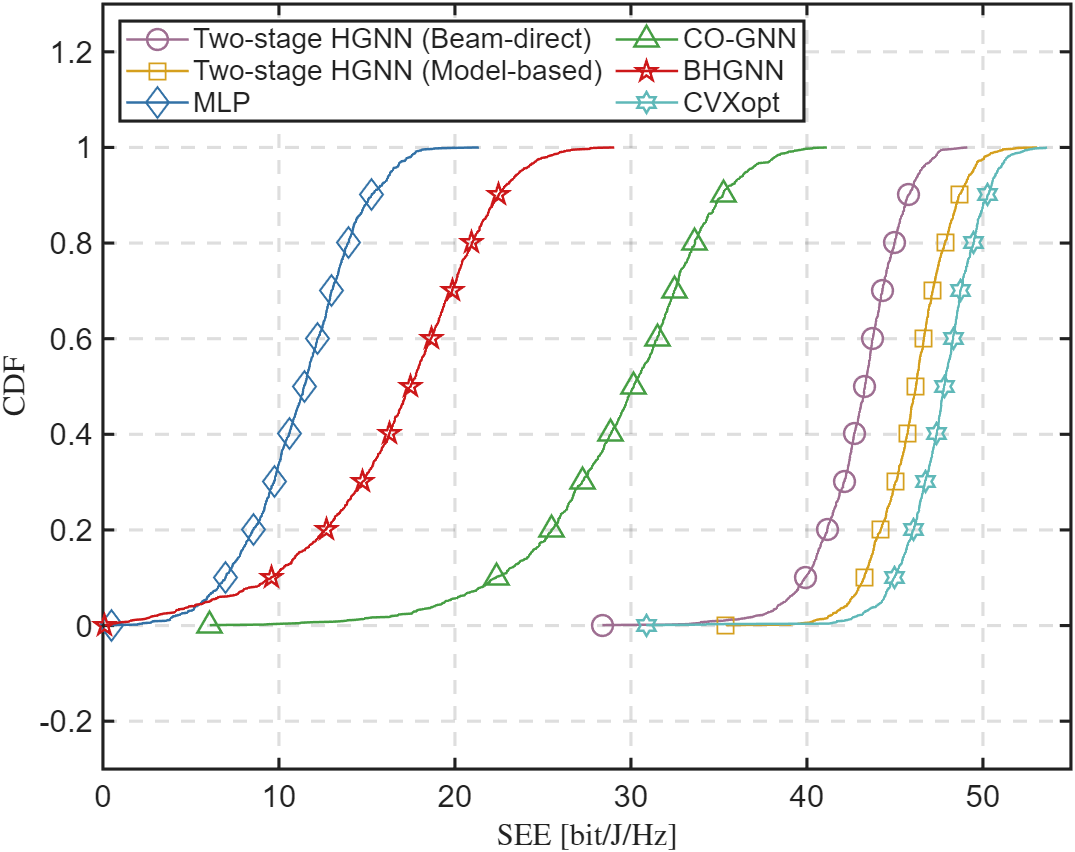} 
  \caption{CDF of SEE for DL-based and CVXopt-based approaches.}
  \label{fig:cdf}
\end{figure}

Fig. \ref{fig:cdf} plots the cumulative distribution function (CDF) of the achievable SEE for the proposed model and baselines as well as the CVXopt-based approach over $1000$ test samples, with $(L_{\rm Tr},K_{\rm Tr},M_{\rm Tr})=(L_{\rm Te},K_{\rm Te},M_{\rm Te})=(8,3,4)$. Compared with the baselines, the two-stage HGNN yields more stable SEE performance and significantly reduces the probability of low SEE.  Particularly, the model-based method allows the two-stage HGNN to achieve performance comparable to that of the CVXopt-based approach for all test samples, while requiring much less inference time.

\begin{figure}[!t]
  \centering
  \includegraphics[width=\columnwidth]{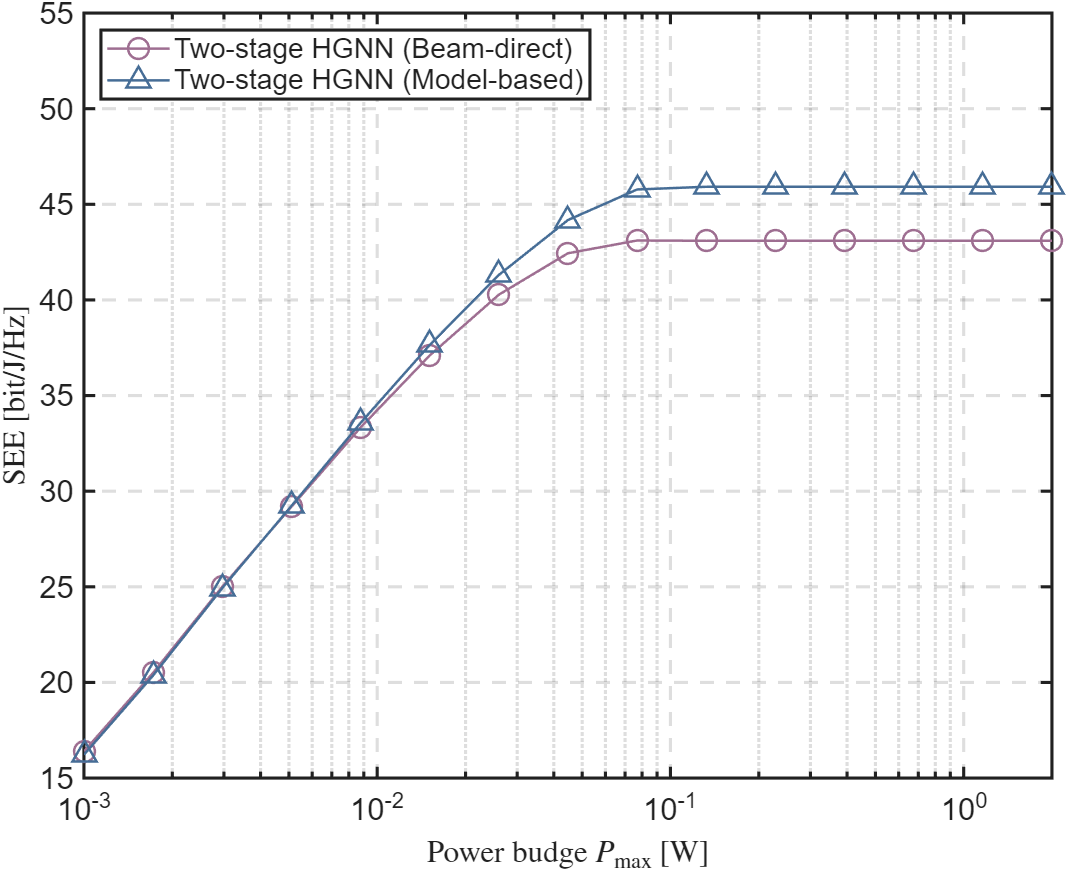} 
  \caption{SEE versus power budget by the two-stage HGNN trained with $P_{\rm max}=1$ W.}
  \label{fig:p_max_performance}
\end{figure}

Fig. \ref{fig:p_max_performance} illustrates the achievable SEE by the proposed two-stage HGNN trained with $P_{\max}=30$ dBm  over the power budget range $[0,33]$ dBm. Results for both the beam-direct and model-based output methods are provided. It can be observed that the SEE first increases and then remains unchanged with the power budget, which is consistent with the theoretical results \cite{li2013energy}. That is, the two-stage HGNN can learn the key characteristics of the SEE function, enabling it to adapt to various power budgets. Such a capability is derived by the parameter-free scaling operation \eqref{af} and \eqref{af2}. Moreover, the model-based output method behaves comparably to the beam-direct one at low power budgets, while achieving superior performance at high power budgets.

\begin{figure}[!t]
  \centering
  \includegraphics[width=\columnwidth]{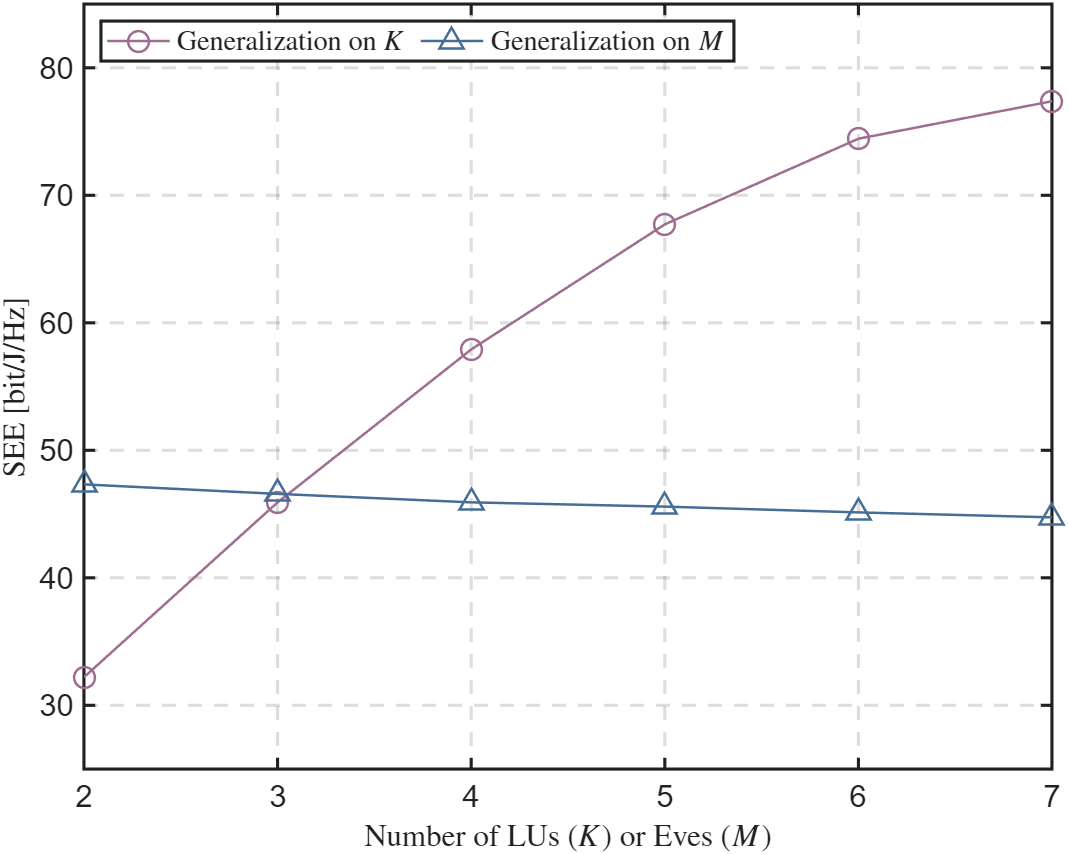} 
  \caption{Generalization performance of the two-stage GNN with the model-based output method with $K_{\rm Tr}=3$ and $M_{\rm Tr}=4$.}
  \label{fig:km_performance}
\end{figure}

Fig. \ref{fig:km_performance} illustrates the scalability of the two-stage HGNN with respect to the number of LUs $K$ or Eves $M$ over a range $[2, 7]$. During the training phase, $K$ and $M$ are set to $3$ and $4$, respectively, and the model-based output method is employed. It can be observed that the two-stage HGNN accurately captures the trend that the achievable SEE increases with $K$, which is consistent with the simulation results in \cite{lu2023secrecy}. Moreover,  as $M$ increases, the SEE achieved by the two-stage HGNN degrades, since it becomes increasingly challenging to prevent information leakage, which is consistent with the simulation results in \cite{Hao2024Robust}. Notably, such scalability is obtained without retraining, making it highly crucial for adapting to the dynamics of wireless networks.

\begin{table*}[ht]    
\belowrulesep=-0.5pt
\aboverulesep=-0.5pt
    \centering  
    \caption{Ablation experiment with $(L_{\rm r},K_{\rm Tr},M_{\rm Tr})=(8,3,4)$.}
    \label{table:Ablation}
    \begin{threeparttable}
    \begin{tabular}{ccc|c c c c c c c|c}  
\toprule
        \multirow{2}{*}{Res} & \multirow{2}{*}{Two-Stage} & \multirow{2}{*}{Model-based} & \multicolumn{7}{c|}{{$(L_{\rm Te},K_{\rm Te},M_{\rm Te})$}} & Average \\
        \cmidrule(lr){4-10}
        & & & $(8,3,4)$  & $(8,2,4)$ & $(8,4,4)$ & $(7,3,4)$ & $(9,3,4)$ & $(8,3,3)$ & $(8,3,5)$ & Gain  \\
        %\cline{4-11}\\
        \hline
        \hline
        $\times$ & $\times$ &$\times$ &  58.70\% & 63.99\% & 36.00\% & 45.54\% & 51.20\% & 58.93\% & 57.14\% & -- \\
        \cline{4-11}
        $\checkmark$& $\times$ & $\times$ & 60.35\% & 55.87\% & 34.69\% &54.80\% & 56.73\% & 60.77\% & 59.10\% & 1.54\% \\
        \cline{4-11}
        $\times$ & $\checkmark$ &$\times$ & 87.12\% & 65.54\% & 38.04\% &86.07\% & 86.21\% & 88.09\% & 53.97\% & 19.07\% \\
        \cline{4-11}
        $\checkmark$ & $\checkmark$ & $\times$ & 90.76\% & 82.01\% & 42.20\% & 89.54\% & 89.73\% & 91.56\% & 79.39\% & 27.67\% \\
        \cline{4-11}
        $\checkmark$ & $\checkmark$ &$\checkmark$ & 96.60\% & 96.80\% & 96.99\% & 96.46\% & 96.43\% & 97.27\% & 95.93\% & 43.56\% \\
        \bottomrule
    \end{tabular} 
    \end{threeparttable}  
\end{table*}  

\subsection{Ablation Experiment}%RIS,user,Eve

Table \ref{table:Ablation} presents the ablation study{\footnote{The ablation study on HGNN is intractable, since the considered system deviates significantly from the homogeneous graph representation.}} to validate the effectiveness of three key components, i.e., the residual connection, the two-stage framework, and the model-based output method. As observed, all components contribute to the performance improvement. For clarity, we report the average gain compared with the model without the three components. Specifically, the residual connection and  the two-stage framework improve the achievable SEE by $1.54\%$ and $19.07\%$, respectively, and their combination yields a $27.67\%$ performance gain. This result demonstrates that the basic GNN model for RIS-assisted systems can be effectively built upon the two-stage framework with residual connections. The model-based output method contributes to an average gain of $43.56\%$, which mainly benefits from its stable performance when scaling to different system configurations. Thus, it is essential to integrate theoretical results into the neural network architecture.

%\textcolor{blue}{, and indicates that compared with the single-stage framework, this two-stage framework has lower fitting difficulty and achieves better performance.}

\section{Conclusion}

This paper focused on developing a GNN-based learning-to-optimize approach for RIS-aided PLS systems. We proposed to jointly exploit the multi-stage framework and the heterogeneous graph representation to tackle the deeply coupled optimization variables and inherent system heterogeneity. Specifically, a two‑stage HGNN is developed to maximize the system SEE. The two-stage HGNN follows a stage-wise graph modeling strategy: the first stage constructs a bipartite graph to learn the optimal RIS phase shift matrix, while the second stage employs a fully-connected heterogeneous graph to generate beamforming and AN vectors. Multi‑head attention mechanisms and residual connections are integrated into both stages to enhance feature learning and model stability. Trained in an unsupervised manner, the two-stage HGNN avoids the dependence on large labeled datasets and reduces the computational burden. Its scalable structure enables flexible adaptation to dynamic numbers of RIS elements, LUs, and Eves. Numerical results confirmed the superiority of the proposed HGNN over state-of-the-art GNNs in RIS-aided PLS scenarios. Compared with CVXopt-based algorithms, it drastically reduces the inference time while limiting the performance degradation to below $4\%$. Furthermore, its scalability under various system configurations is verified.

%In summary, this work provides an efficient GNN-based solution for PLS optimization in RIS-assisted MISO systems, addressing the performance-computational complexity trade-off. Future work may extend the framework to complex scenarios (e.g., multi-RIS, mobile users) and enhance its robustness to dynamic channels.

    \appendix

    \subsection{Edge-Based Graph Attention Operator}

Given a graph $\cal G$ and the corresponding node feature matrices $\{{\bf X}_n\}$, the edge-based graph attention operator is detailed as follows to update the node feature matrices.

For clarity, we define the ${\bf x}_i$ (derived  from $\{{\bf X}_n\}$) by the node feature of the $i$-th node, ${\cal N}(i)$ by the neighboring source node set of the $i$-th node, and ${\bf y}_{i,i^{\prime}}$ by the directed edge feature from the $i^{\prime}$-th ($i^{\prime}\in{\cal N}(i) $) node to the $i$-th node, which keeps unchanged.

The $i$-th target node assigns an attention coefficient associated with the $d$-th  ($d\in\{1,2,\ldots,D\}$) head to its $j$-th ($j\in{\cal N}(i) $) source node, which is calculated by \eqref{eq:attn_with_edge}, where ${\bf a}_d$ denotes the learnable attention weight vector, and ${\bf W}_d^{({\rm S)}}$, ${\bf W}_d^{({\rm N)}}$, and ${\bf W}_d^{({\rm E)}}$ denote the learnable  projection matrices for target node, source node, and edge features, respectively.

\begin{figure*}    \begin{flalign}\label{eq:attn_with_edge}
\alpha^{(i,j)}_{d} = \frac{\expf\left( \mathbf{a}_{d}^T {\rm LeakyReLU}\left( \mathbf{W}_{d}^{(\text{S})} \mathbf{X}_i + \mathbf{W}_{d}^{(\text{N})} \mathbf{x}_j + \mathbf{W}_{d}^{(\text{E})} \mathbf{y}_{i,j} \right) \right)}{\sum\nolimits_{i^{\prime} \in {\cal N}  \left(i\right)} \exp \left( \mathbf{a}_{d}^{T} {\rm LeakyReLU}\left( \mathbf{W}_{d}^{(\text{S})} \mathbf{x}_i + \mathbf{W}_{d}^{(\text{N})} \mathbf{x}_{i^{\prime}} + \mathbf{W}_{d}^{(\text{E})} \mathbf{y}_{i,i^{\prime}} \right) \right)}
\end{flalign}
\hrule
\end{figure*}

    Then, the $i$-th node aggregates the node features of itself and its neighboring target nodes based on the $d$-th attention head to obtain  
\begin{align}\label{eq:agg_with_edge}
{{\bm \beta}}^{(i)}_{d} = \mathbf{W}_{d}^{(\text{S})} \mathbf{x}_i + \sum\nolimits_{i^{\prime} \in {\cal N}\left({i}\right)} \alpha^{(i,i^{\prime})}_{d} \left( \mathbf{W}_d^{(\text{N})} \mathbf{x}_{i^{\prime}} + \mathbf{W}_{d}^{(\text{E})} \mathbf{y}_{i,i^{\prime}} \right),
\end{align}

By concatenating the $D$ aggregated features, the feature of the $i$-th node is updated to $\widetilde{{\bf x}}_{i}$, which is given by 
\begin{flalign}\label{eq:final_with_edge}
\widetilde{{\bf x}}_{i}  =  \Concat\left( \left\{ \bm{\beta}_{d}^{(i)} \right\}\right).
\end{flalign}

    \subsection{Edge-Free Graph Attention Operator}

    Similarly, the edge-free graph attention operator updates node features of a given graph $\cal G$ without involving the edge features. 

    We reuse the symbols defined for simplicity. The $d$-th attention coefficient assigned by the $i$-th target node to the $j$-th source node is given by
\begin{flalign}\label{eq:attn_without_edge}
&\alpha^{(i,j)}_{d} = \\
&\frac{\expf\left( \mathbf{a}_{d}^T {\rm LeakyReLU}\left( \mathbf{W}_{d}^{(\text{S})} \mathbf{X}_i + \mathbf{W}_{d}^{(\text{N})} \mathbf{x}_j \right) \right)}{\sum\nolimits_{i^{\prime} \in {\cal N}  \left(i\right)} \exp \left( \mathbf{a}_{d}^{T} {\rm LeakyReLU}\left( \mathbf{W}_{d}^{(\text{S})} \mathbf{x}_i + \mathbf{W}_{d}^{(\text{N})} \mathbf{x}_{i^{\prime}} \right) \right)}.\nonumber
\end{flalign}

Then, the node feature of the $i$-th node is updated by
\begin{flalign}\label{eq:final_without_edge}
\widetilde{{\bf x}}_{i}  =  \Concat\left(\left\{\sum\nolimits_{i^{\prime} \in {\cal N}\left({i}\right)} \alpha^{(i,i^{\prime})}_{d} \mathbf{W}_d^{(\text{N})} \mathbf{x}_{i^{\prime}}  \right\}\right).
\end{flalign}

\subsection{Hybrid ZF and MRT method}

 We represent ${\bf w}_k$ and ${\bf z}_m$ as 
\begin{flalign}
    {\bf w}_k &= \sqrt{p_k}{\overline{\bf w}_k}\left(\alpha_k \right),~\|{\overline{\bf w}_k}\left(\alpha_k \right)\|=1,\\
    {\bf z}_m &= \sqrt{p_m}{\overline{\bf z}_m}\left(\beta_m \right),~\|{\overline{\bf z}_m}\left(\beta_m \right)\|=1.
\end{flalign}

For ${\bf w}_k$, the hybrid ZF and MRT scheme is adopted to design its direction, which is given by
\begin{flalign}
\overline {\bf w}_k\left(\alpha_k \right) =\frac{\alpha_k\frac{{\bf v}_k}{\left\|{\bf v}_k\right\|} + \left(1-\alpha_k\right)\frac{\widetilde{\bf h}_k\left({\bm\Phi}\right)}{\left\|\widetilde{\bf h}_k\left({\bm\Phi}\right)\right\|}}{\left\|{\alpha_k\frac{{\bf v}_k}{\left\|{\bf v}_k\right\|} + \left(1-\alpha_k\right)\frac{\widetilde{\bf h}_k\left({\bm\Phi}\right)}{\left\|\widetilde{\bf h}_k\left({\bm\Phi}\right)\right\|}}\right\|},
\end{flalign}
where $\alpha_k \in [0,1]$ is a learnable parameter, and ${\bf{v}}_k$ is the $k$-th column of 
\begin{flalign}
    {\bf V} \triangleq {\bf G}^H\left({\bf G}{\bf G}^H\right)^{-1},
\end{flalign}
where ${\bf G} \triangleq [\widetilde{\bf h}_1\left({\bm\Phi}\right),\widetilde{\bf h}_2\left({\bm\Phi}\right),\ldots,\widetilde{\bf h}_K\left({\bm\Phi}\right)]$.

For ${\bf z}_m$, its direction is given by
\begin{flalign}
\overline {\bf z}_m\left(\beta_m \right) = \frac{\beta_m\frac{{\bf v}_{{\rm{Eve}},m}}{\left\|{\bf v}_{{\rm{Eve}},m}\right\|} + \left(1-\beta_m\right)\frac{\widetilde{\bf f}_{m}\left({\bm\Phi}\right)}{\left\|\widetilde{\bf f}_{m}\left({\bm\Phi}\right)\right\|}}{\left\| {\beta_m\frac{{\bf v}_{{\rm{Eve}},m}}{\left\|{\bf v}_{{\rm{Eve}},m}\right\|} + \left(1-\beta_m\right)\frac{\widetilde{\bf f}_{m}\left({\bm\Phi}\right)}{\left\|\widetilde{\bf f}_{m}\left({\bm\Phi}\right)\right\|}}\right\|},
\end{flalign}
where $\beta_m \in [0,1]$ is a learnable parameter, and ${\bf{v}}_{{\rm{Eve}},m}$ is the $(k+1)$-th column of 
\begin{flalign}
    {\bf V}_{{\rm{Eve}},m} \triangleq {\bf G}_{{\rm{Eve}},m}^H\left({\bf G}_{{\rm{Eve}},m}{\bf G}_{{\rm{Eve}},m}^H\right)^{-1},
\end{flalign}
where ${\bf G}_{{\rm{Eve}},m} \triangleq [\widetilde{\bf h}_1\left({\bm\Phi}\right),\widetilde{\bf h}_2\left({\bm\Phi}\right),\ldots,\widetilde{\bf h}_K\left({\bm\Phi}\right),\widetilde{\bf f}_m \left({\bm\Phi}\right)]$.

\bibliographystyle{IEEEtran} 
\bibliography{refs}

@ARTICLE{bck1,
  author={Wu, Qingqing and Zhang, Rui},
  journal={IEEE Commun. Mag.}, 
  title={Towards Smart and Reconfigurable Environment: Intelligent Reflecting Surface Aided Wireless Network}, 
  year={2020},
  month={Jan.},
  volume={58},
  number={1},
  pages={106-112},
  doi={10.1109/MCOM.001.1900107}}

@ARTICLE{bck2,
  author={ElMossallamy, Mohamed A. and Zhang, Hongliang and Song, Lingyang and Seddik, Karim G. and Han, Zhu and Li, Geoffrey Ye},
  journal={IEEE Trans. Cognit. Commun. Networking}, 
  title={Reconfigurable Intelligent Surfaces for Wireless Communications: Principles, Challenges, and Opportunities}, 
  year={2020},
  month={Sept.},
  volume={6},
  number={3},
  pages={990-1002},
  doi={10.1109/TCCN.2020.2992604}}

@ARTICLE{bck3,
  author={Liu, Yuanwei and Liu, Xiao and Mu, Xidong and Hou, Tianwei and Xu, Jiaqi and Di Renzo, Marco and Al-Dhahir, Naofal},
  journal={IEEE Commun. Surv. Tutorials}, 
  title={Reconfigurable Intelligent Surfaces: Principles and Opportunities}, 
  year={2021},
  volume={23},
  number={3},
  pages={1546-1577},
  doi={10.1109/COMST.2021.3077737}}

@ARTICLE{hong_tcom_2020,
  author={Hong, Sheng and Pan, Cunhua and Ren, Hong and Wang, Kezhi and Nallanathan, Arumugam},
  journal={IEEE Trans. Commun.},
  title={Artificial-Noise-Aided Secure {MIMO} Wireless Communications via Intelligent Reflecting Surface}, 
  year={2020},
  month={Dec.},
  volume={68},
  number={12},
  pages={7851-7866},
  doi={10.1109/TCOMM.2020.3024621}}

@ARTICLE{Guo_twc_2020,
  author={Guo, Huayan and Liang, Ying-Chang and Chen, Jie and Larsson, Erik G.},
  journal={IEEE Trans. Wireless Commun.}, 
  title={Weighted Sum-Rate Maximization for Reconfigurable Intelligent Surface Aided Wireless Networks}, 
  year={2020},
  month={May},
  volume={19},
  number={5},
  pages={3064-3076},
  doi={10.1109/TWC.2020.2970061}}

@ARTICLE{huang_jsac_2020,
  author={Huang, Chongwen and Mo, Ronghong and Yuen, Chau},
  journal={IEEE J. Sel. Areas Commun.	}, 
  title={Reconfigurable Intelligent Surface Assisted Multiuser {MISO} Systems Exploiting Deep Reinforcement Learning}, 
  year={2020},
  month={Aug.},
  volume={38},
  number={8},
  pages={1839-1850},
  doi={10.1109/JSAC.2020.3000835}}

@ARTICLE{lu2023secrecy,
  author={Lu, Yang},
  journal={IEEE Trans. Veh. Technol.}, 
  title={Secrecy Energy Efficiency in {RIS}-Assisted Networks}, 
  year={2023},
  month={Sept.},
  volume={72},
  number={9},
  pages={12419-12424},
  doi={10.1109/TVT.2023.3269905}}

@ARTICLE{sun_tsp_2018,
  author={Sun, Haoran and Chen, Xiangyi and Shi, Qingjiang and Hong, Mingyi and Fu, Xiao and Sidiropoulos, Nicholas D.},
  journal={IEEE Trans. Signal Process.}, 
  title={Learning to Optimize: Training Deep Neural Networks for Interference Management}, 
  year={2018},
  month={Oct.},
  volume={66},
  number={20},
  pages={5438-5453},
  doi={10.1109/TSP.2018.2866382}}

@ARTICLE{gnn1,
  author={Lu, Yang and Zhang, Shengli and Liu, Chang and Zhang, Ruichen and Ai, Bo and Niyato, Dusit and Ni, Wei and Wang, Xianbin and Jamalipour, Abbas},
  journal={IEEE Commun. Surv. Tutorials}, 
  title={Agentic Graph Neural Networks for Wireless Communications and Networking Toward Edge General Intelligence: A Survey}, 
  year={2026},
  volume={28},
  number={},
  pages={4519-4554},
  doi={10.1109/COMST.2026.3651990}}

@ARTICLE{gnn2,
  author={Lu, Yang and Li, Yuhang and Zhang, Ruichen and Chen, Wei and Ai, Bo and Niyato, Dusit},
  journal={IEEE Wireless Commun.}, 
  title={Graph Neural Networks for Wireless Networks: Graph Representation, Architecture and Evaluation}, 
  year={2025},
  month={Feb.},
  volume={32},
  number={1},
  pages={150-156},
  doi={10.1109/MWC.006.2400131}}

@ARTICLE{Song_tvt_2024,
  author={Song, Zihan and Lu, Yang and Chen, Xianhao and Ai, Bo and Zhong, Zhangdui and Niyato, Dusit},
  journal={IEEE Trans. Veh. Technol.}, 
  title={A Deep Learning Framework for Physical-Layer Secure Beamforming}, 
  year={2024},
  month={Dec.},
  volume={73},
  number={12},
  pages={19844-19849},
  doi={10.1109/TVT.2024.3442167}}

@ARTICLE{li_twc_2024,
  author={Li, Yuhang and Lu, Yang and Ai, Bo and Dobre, Octavia A. and Ding, Zhiguo and Niyato, Dusit},
  journal={IEEE Trans. Wireless Commun.	}, 
  title={{GNN}-Based Beamforming for Sum-Rate Maximization in {MU-MISO} Networks}, 
  year={2024},
  month={Aug.},
  volume={23},
  number={8},
  pages={9251-9264},
  doi={10.1109/TWC.2024.3361174}}

@ARTICLE{li2013energy,
  author={Li, Yuzhou and Sheng, Min and Yang, Chungang and Wang, Xijun},
  journal={IEEE Commun. Lett.	}, 
  title={Energy Efficiency and Spectral Efficiency Tradeoff in Interference-Limited Wireless Networks}, 
  year={2013},
  month={Oct.},
  volume={17},
  number={10},
  pages={1924-1927},
  doi={10.1109/LCOMM.2013.082613.131286}}

@ARTICLE{shen_twc_2023,
  author={Shen, Yifei and Zhang, Jun and Song, S. H. and Letaief, Khaled B.},
  journal={IEEE Trans. Wireless Commun.}, 
  title={Graph Neural Networks for Wireless Communications: From Theory to Practice}, 
  year={2023},
  month={May},
  volume={22},
  number={5},
  pages={3554-3569},
  doi={10.1109/TWC.2022.3219840}}

@INPROCEEDINGS{Abdalla_ICC_2025,
  author={Abdalla, Aly Sabri and Marojevic, Vuk},
  booktitle={Proc. ICC}, 
  title={Enhancing Secrecy Energy Efficiency in {RIS}-Aided Aerial Mobile Edge Computing Networks: A Deep Reinforcement Learning Approach}, 
  year={2025},
  volume={},
  number={},
  pages={6585-6590}
}

@ARTICLE{Yang2024GATPrecoding,
  author={Yang, Junjie and Xu, Jie and Zhang, Yinghui and Zheng, Hao and Zhang, Tiankui},
  journal={IEEE Trans. Veh. Technol.}, 
  title={Graph Attention Network-Based Precoding for Reconfigurable Intelligent Surfaces Aided Wireless Communication Systems}, 
  year={2024},
  month={June},
  volume={73},
  number={6},
  pages={9098-9102},
  doi={10.1109/TVT.2024.3354967}}

@ARTICLE{he2025ris,
author = {Changpeng He and Yang Lu and Yanqing Xu and Chong-Yung Chi and Bo Ai and Arumugam Nallanathan},
title = {{RIS}-Assisted Downlink Pinching-Antenna Systems: {GNN}-Enabled Optimization Approaches},
journal = {arXiv preprint arXiv:2511.20305},
year = {2025},
month = {Nov.},
eprint = {2511.20305},
archivePrefix = {arXiv},
primaryClass = {cs.NI},
doi = {10.48550/arXiv.2511.20305},
url = {https://arxiv.org/abs/2511.20305}}

@ARTICLE{yeh2024enhanced,
  author={Yeh, Ting-Ju and Tsai, Wen-Chiao and Chen, Chi-Wei and Wu, An-Yeu},
  journal={IEEE Commun. Lett.}, 
  title={Enhanced-{GNN} With Angular CSI for Beamforming Design in {IRS}-Assisted mm{Wave} Communication Systems}, 
  year={2024},
  month={April},
  volume={28},
  number={4},
  pages={827-831},
  doi={10.1109/LCOMM.2024.3363450}}

@ARTICLE{tang2025joint,
  author={Tang, Huijun and Zhang, Jieling and Zhao, Zhidong and Wu, Huaming and Sun, Hongjian and Jiao, Pengfei},
  journal={IEEE Trans. Wireless Commun.}, 
  title={Joint Optimization Based on Two-Phase {GNN} in {RIS}- and {DF}-Assisted {MISO} Systems With Fine-Grained Rate Demands}, 
  year={2025},
  month={Dec.},
  volume={24},
  number={12},
  pages={9989-10002},
  doi={10.1109/TWC.2025.3576298}}

@ARTICLE{liu2025beamforming,
  author={Liu, Mengbing and Huang, Chongwen and Alhammadi, Ahmed and Di Renzo, Marco and Debbah, Mérouane and Yuen, Chau},
  journal={IEEE Trans. Wireless Commun.	}, 
  title={Beamforming Design and Association Scheme for Multi-{RIS} Multi-User mm{Wave} Systems Through Graph Neural Networks}, 
  year={2025},
  month={Sept.},
  volume={24},
  number={9},
  pages={7940-7954},
  doi={10.1109/TWC.2025.3563529}}

@ARTICLE{le2025graph,
  author={Le, Ha An and Van Chien, Trinh and Choi, Wan},
  journal={IEEE Trans. Commun.}, 
  title={Graph Neural Network-Based Active and Passive Beamforming for Distributed {STAR}-{RIS}-Assisted Multi-User {MISO} Systems}, 
  year={2025},
  month={Oct.},
  volume={73},
  number={10},
  pages={9299-9312},
  doi={10.1109/TCOMM.2025.3558504}}

@ARTICLE{Zhang2026Secrecy,
  author={Zhang, Jieling and Tang, Huijun and Jiao, Pengfei and Wu, Huaming and Zhao, Zhidong and Li, Ruidong},
  journal={IEEE Internet Things J.}, 
  title={Secrecy Rate Optimization based on {GNN} for {RIS}-Assisted {ISAC} System}, 
  year={2026},
  volume={},
  number={},
  pages={1-1},
  doi={10.1109/JIOT.2026.3664707}}

@ARTICLE{liang2025heterogeneous,
  author={Liang, Linlin and Tian, Zongkai and Huang, Haiyan and Li, Xiaoyan and Yin, Zhisheng and Zhang, Dehua and Zhang, Nina and Zhai, Wenchao},
  journal={IEEE Internet Things J.}, 
  title={Heterogeneous Secure Transmissions in {IRS}-Assisted {NOMA} Communications: {CO}-{GNN} Approach}, 
  year={2025},
  month={Aug.},
  volume={12},
  number={16},
  pages={34113-34125},
  doi={10.1109/JIOT.2025.3577332}}

@ARTICLE{Hao2024Robust,
  author={Hao, Wanming and Li, Junjie and Sun, Gangcan and Huang, Chongwen and Zeng, Ming and Dobre, Octavia A. and Yuen, Chau},
  journal={IEEE Trans. Commun.}, 
  title={Robust Security Energy Efficiency Optimization for {RIS}-Aided Cell-Free Networks With Multiple Eavesdroppers}, 
  year={2024},
  month={Dec.},
  volume={72},
  number={12},
  pages={7401-7416},
  doi={10.1109/TCOMM.2024.3411769}}

@ARTICLE{pan_jstsp_2022,
  author={Pan, Cunhua and Zhou, Gui and Zhi, Kangda and Hong, Sheng and Wu, Tuo and Pan, Yijin and Ren, Hong and Renzo, Marco Di and Lee Swindlehurst, A. and Zhang, Rui and Zhang, Angela Yingjun},
  journal={IEEE J. Sel. Top. Signal Process.}, 
  title={An Overview of Signal Processing Techniques for {RIS}/{IRS}-Aided Wireless Systems}, 
  year={2022},
  month={Aug.},
  volume={16},
  number={5},
  pages={883-917},
  doi={10.1109/JSTSP.2022.3195671}}

@ARTICLE{Zhang_jsac_2023,
  author={Zhang, Ruichen and Xiong, Ke and Lu, Yang and Fan, Pingyi and Ng, Derrick Wing Kwan and Letaief, Khaled B.},
  journal={IEEE J. Sel. Areas Commun.	}, 
  title={Energy Efficiency Maximization in {RIS}-Assisted {SWIPT} Networks With {RSMA}: A {PPO}-Based Approach}, 
  year={2023},
  month={May},
  volume={41},
  number={5},
  pages={1413-1430},
  doi={10.1109/JSAC.2023.3240707}}

@ARTICLE{S2025Optimization,
  author={Nguyen, Quynh-Suong and Dang, Xuan-Toan and Shin, Oh-Soon},
  journal={IEEE Trans. Veh. Technol.}, 
  title={Optimization of {RIS}-assisted Cell-free Massive {MIMO} Systems with Heterogeneous Graph Neural Networks under Imperfect Channel Estimation}, 
  year={2025},
  volume={},
  number={},
  pages={1-16},
  doi={10.1109/TVT.2025.3624845}}

@ARTICLE{Yin2026Graph,
  author={Yin, Bo and Schampheleer, Jorn and Joseph, Wout and Deruyck, Margot},
  journal={IEEE Trans.  Wireless Commun.}, 
  title={Graph Neural Network-Based Energy-Efficient Optimization for {RIS}-Assisted Wireless Networks}, 
  year={2026},
  volume={25},
  number={},
  pages={664-680},
  doi={10.1109/TWC.2025.3585772}}

@INPROCEEDINGS{GCN,
  author    = {Kipf, Thomas N. and Welling, Max},
  title     = {Semi-Supervised Classification with Graph Convolutional Networks},
  booktitle = {Proc. ICLR},
  year      = {2017},
  month     = apr
}

@INPROCEEDINGS{GAT,
  author    = {Veli{\v{c}}kovi{\'{c}}, Petar and Cucurull, Guillem and Casanova, Arantxa and Romero, Adriana and Li{\`{o}}, Pietro and Bengio, Yoshua},
  title     = {Graph Attention Networks},
  booktitle = {Proc. ICLR},
  year      = {2018},
  month     = apr
}

@ARTICLE{Zheng2022Residual,
  author    = {Zheng, Xuhui and Liu, Ziyan and Liang, Jing and Wu, Yingyu and Chen, Yunlei and Zhang, Qian},
  title     = {Residual Learning and Multi-Path Feature Fusion-Based Channel Estimation for Millimeter-Wave Massive {MIMO} System},
  journal   = {Entropy},
  year      = {2022},
  volume    = {24},
  number    = {2},
  pages     = {292},
  month     = feb,
  doi       = {10.3390/e24020292},
  publisher = {MDPI}
}

@INPROCEEDINGS{wang2019heterogeneous,
  author    = {Wang, Xiao and Ji, Houye and Shi, Chuan and Wang, Bai and Ye, Yanfang and Cui, Peng and Yu, Philip S.},
  title     = {Heterogeneous Graph Attention Network},
  booktitle = {Proc. World Wide Web Conf. (WWW)},
  year      = {2019},
  pages     = {2022--2032},
  month     = may
}

@ARTICLE{Saikia2023Beamforming,
  author={Saikia, Prajwalita and Singh, Keshav and Singh, Sandeep Kumar and Huang, Wan-Jen and Li, Chih-Peng and Biswas, Sudip},
  journal={IEEE Access}, 
  title={Beamforming Design in Vehicular Communication Systems With Multiple Reconfigurable Intelligent Surfaces: A Deep Learning Approach}, 
  year={2023},
  volume={11},
  number={},
  pages={100832-100844},
  doi={10.1109/ACCESS.2023.3314668}}

@ARTICLE{Gao2020Unsupervised,
  author={Gao, Jiabao and Zhong, Caijun and Chen, Xiaoming and Lin, Hai and Zhang, Zhaoyang},
  journal={IEEE Commun. Lett.}, 
  title={Unsupervised Learning for Passive Beamforming}, 
  year={2020},
  month={May},
  volume={24},
  number={5},
  pages={1052-1056},
  doi={10.1109/LCOMM.2020.2965532}}

@ARTICLE{Ma2026Unsupervised,
  author={Ma, Yu and Zhou, Xingyu and Li, Xiao and Liang, Le and Jin, Shi},
  journal={IEEE Trans. Cognit. Commun. Networking}, 
  title={Unsupervised Learning-Based Joint Resource Allocation and Beamforming Design for {RIS}-Assisted {MISO}-{OFDMA} Systems}, 
  year={2026},
  volume={12},
  number={},
  pages={2251-2264},
  doi={10.1109/TCCN.2025.3592931}}

@ARTICLE{zhang2023Passive,
  author  = {Zhang, Hui and Jia, Qiming and Li, Meikun and Wang, Jingjing and Song, Yuxin},
  title   = {{Passive Beamforming Design of {IRS}-Assisted {MIMO} Systems Based on Deep Learning}},
  journal = {Sensors},
  year    = {2023},
  volume  = {23},
  number  = {16},
  pages   = {7164},
  month   = aug,
  doi     = {10.3390/s23167164},
  issn    = {1424-8220},
  publisher = {MDPI}
}

@ARTICLE{he2025ICGNN,
  author={He, Changpeng and Lu, Yang and Ai, Bo and Dobre, Octavia A. and Ding, Zhiguo and Niyato, Dusit},
  journal={IEEE Trans. Mob. Comput.}, 
  title={{ICGNN}: Graph Neural Network Enabled Scalable Beamforming for {MISO} Interference Channels}, 
  year={2025},
  month={Oct.},
  volume={24},
  number={10},
  pages={10778-10791},
  doi={10.1109/TMC.2025.3570648}}

@ARTICLE{Amiriara2023IRS-User,
  author={Amiriara, Hamid and Ashtiani, Farid and Mirmohseni, Mahtab and Nasiri-Kenari, Masoumeh},
  journal={IEEE Trans. Veh. Technol.}, 
  title={{IRS}-User Association in {IRS}-Aided {MISO} Wireless Networks: Convex Optimization and Machine Learning Approaches}, 
  year={2023},
  month={Nov.},
  volume={72},
  number={11},
  pages={14305-14316},
  doi={10.1109/TVT.2023.3282272}}

@ARTICLE{10851843,
  author={Zhang, Rongsheng and Lu, Yang and Chen, Wei and Ai, Bo and Ding, Zhiguo},
  journal={IEEE Trans. Wireless Commun.}, 
  title={Model-Based {GNN} Enabled Energy-Efficient Beamforming for Ultra-Dense Wireless Networks}, 
  year={2025},
  volume={24},
  number={4},
  pages={3333-3345},
  doi={10.1109/TWC.2025.3530003}}

\end{document}